\documentclass[twocolumn,twocolappendix]{aastex63}

\usepackage{graphicx}
\usepackage{amsmath}
\usepackage{amssymb}
\usepackage{bm}
\usepackage{multirow}
\usepackage{upgreek}

\received{June 1, 2019}
\revised{January 10, 2019}
\accepted{\today}
\shorttitle{FRB 121102 RM Evolution}
\shortauthors{Hilmarsson et al.}
\graphicspath{{./}}

\newcommand{\frb}{FRB~121102}
\newcommand{\magn}{PSR J1745$-$2900}

\begin{document}

\title{Rotation Measure Evolution of the Repeating Fast Radio Burst Source \frb}

\correspondingauthor{G.~H.~Hilmarsson}
\email{henning@mpifr-bonn.mpg.de}

\author{G.~H.~Hilmarsson}
\affiliation{Max-Planck-Institute f{\"u}r Radioastronomie,
Auf dem H{\"u}gel 69, D-53121, 
Bonn, Germany}

\author{D.~Michilli}
\affiliation{Department of Physics, McGill University, 3600 rue University, Montr\'eal, QC H3A 2T8, Canada}
\affiliation{McGill Space Institute, McGill University, 3550 rue University, Montr\'eal, QC H3A 2A7, Canada}

\author{L.~G.~Spitler}
\affiliation{Max-Planck-Institute f{\"u}r Radioastronomie,
Auf dem H{\"u}gel 69, D-53121,
Bonn, Germany}

\author{R.~S.~Wharton}
\affiliation{Max-Planck-Institute f{\"u}r Radioastronomie,
Auf dem H{\"u}gel 69, D-53121,
Bonn, Germany}

\author{P.~Demorest}
\affiliation{National Radio Astronomy Observatory,
Socorro, NM 87801,
USA}

\author{G.~Desvignes}
\affiliation{Max-Planck-Institute f{\"u}r Radioastronomie,
Auf dem H{\"u}gel 69, D-53121,
Bonn, Germany}
\affiliation{Laboratoire d'\'Etudes Spatiales et d'Instrumentation en Astrophysique, Observatoire de Paris, Universit\'e PSL\\ CNRS, Sorbonne Universit\'e, Universit\'e de Paris, 5 place Jules Janssen, 92195 Meudon, France}

\author{K.~Gourdji}
\affiliation{Anton Pannekoek Institute for Astronomy, University of Amsterdam,
Science Park 904, 1098 XH,
Amsterdam, The Netherlands}

\author{S.~Hackstein}
\affiliation{Hamburger Sternwarte, University of Hamburg,
Gojenbergsweg 112, D-21029,
Hamburg, Germany}

\author{J.~W.~T.~Hessels}
\affiliation{ASTRON, Netherlands Institute for Radio Astronomy,
Oude Hoogeveensedijk 4, 7991 PD,
Dwingeloo, The Netherlands}
\affiliation{Anton Pannekoek Institute for Astronomy, University of Amsterdam,
Science Park 904, 1098 XH,
Amsterdam, The Netherlands}

\author{K.~Nimmo}
\affiliation{ASTRON, Netherlands Institute for Radio Astronomy,
Oude Hoogeveensedijk 4, 7991 PD,
Dwingeloo, The Netherlands}
\affiliation{Anton Pannekoek Institute for Astronomy, University of Amsterdam,
Science Park 904, 1098 XH,
Amsterdam, The Netherlands}

\author{A.~D.~Seymour}
\affiliation{Green Bank Observatory,
PO Box 2, WV 24944,
Green Bank, USA}

\author{M.~Kramer}
\affiliation{Max-Planck-Institute f{\"u}r Radioastronomie,
Auf dem H{\"u}gel 69, D-53121,
Bonn, Germany}

\author{R.~Mckinven}
\affiliation{David A. Dunlap Department of Astronomy and Astrophysics, University of Toronto,
50 St. George Street ON M5S 3H4,
Toronto, Canada}
\affiliation{Dunlap Institute for Astronomy \& Astrophysics, University of Toronto,
50 St. George Street, ON M5S 3H4,
Toronto, Canada}












\begin{abstract}
The repeating fast radio burst source \frb\ has been shown to  
have an exceptionally high and variable Faraday rotation measure (RM), 
which must be imparted within its host galaxy and likely 
by or within its 
local environment.  In the redshifted ($z=0.193$) source reference frame, 
the RM decreased from $1.46\times10^5$~rad~m$^{-2}$ to $1.33\times10^5$~rad~m$^{-2}$
between January and August 2017, 
showing day-timescale variations of $\sim200$~rad~m$^{-2}$.
Here we present sixteen \frb\ RMs from burst
detections with the Arecibo 305-m radio telescope, the Effelsberg 100-m, 
and the Karl G. Jansky Very Large Array, providing
a record of \frb's RM over a 2.5-year timespan.
Our observations show a decreasing trend in RM,
although the trend is not linear,
dropping by an average of 15\% year$^{-1}$
and is $\sim9.7\times10^4$~rad~m$^{-2}$
at the most recent epoch of August 2019. 
Erratic, short-term RM variations of $\sim10^3$~rad~m$^{-2}$ week$^{-1}$
were also observed between MJDs 58215--58247.
A decades-old neutron star embedded within a still-compact 
supernova remnant or a neutron star near a massive black hole and its accretion torus
have been proposed to explain the high RMs.
We compare the observed RMs to theoretical models
describing the RM evolution for FRBs originating within a supernova 
remnant. 
\frb's age is unknown, 
and we find that the models
agree for source ages of $\sim6-17$~years 
at the time of the first available RM measurements in 2017.
We also draw comparisons to the decreasing RM of the
Galactic center magnetar, \magn.
\end{abstract}

\keywords{editorials, notices --- 
miscellaneous --- catalogs --- surveys}


\section{introduction}
\label{sect:intro}

Fast radio bursts (FRBs) are millisecond duration radio transients,
whose origins are still unknown \citep{2019A&ARv..27....4P}. 
Of the roughly 100 FRBs published so far\footnote{\url{frbcat.org}} \citep{2016PASA...33...45P}, 
around ten have been localised to a host
galaxy \citep{2017Natur.541...58C,Bannistereaaw5903,
2019Natur.572..352R,2019Sci...366..231P,2020Natur.577..190M,
2020Natur.581..391M}, 
confirming their extragalactic origins.
Some FRBs have also been observed to repeat; the first discovered,
and most observed so far, is
\frb{} \citep{2016Natur.531..202S}, 
and more repeating FRBs have been detected
by the Canadian Hydrogen Intensity Mapping Experiment (CHIME) radio telescope
\citep{2019Natur.566..235C,2019ApJ...885L..24C,2020arXiv200103595F}
and the Australian Square Kilometre Array Pathfinder 
\citep[ASKAP, e.g.][]{2019ApJ...887L..30K}.

Polarisation properties of FRBs can reveal the nature of
their local environment, as well as the FRB emission process and its geometry,
thus adding constraints to progenitor theories.
Polarisation fractions and rotation measures (RMs) have
been determined for 20 FRBs \citep{2016PASA...33...45P}.
Linear polarisation fractions ranging from $\sim 0$ to $\sim 100$\%
have been measured,
and the absolute RM values
are in the range $\sim$10--500 rad m$^{-2}$,
with the exception of \frb{}, which has an
exceptionally high RM of $\sim$10$^5$ rad m$^{-2}$.
\frb{}'s RM has also proven to be highly variable,
with a decrease of $\sim10\%$ between epochs separated
by seven months \citep{2018Natur.553..182M}.
To be able to observe such a high RM, a narrow channel
bandwidth or a high observing frequency are required
in order to avoid intra-channel depolarisation.
Typical pulsar instrumentation has channel bandwidths
of $\sim1$~MHz, so high frequency observations are required
to observe high RMs.

In the original discovery of \frb{}, the dispersion measure (DM)
was found to be $557\pm2$~pc~cm$^{-3}$ \citep{2014ApJ...790..101S}. 
In more recent observations, \frb{} has exhibited
an increase in the measured DM, $560.6\pm0.1$~pc~cm$^{-3}$
at 1.4~GHz
in \citet{2019ApJ...876L..23H} and $563.6\pm0.5$~pc~cm$^{-3}$
at 0.6~GHz
in \citet{2019ApJ...882L..18J}, 
revealing an average increase of roughly 1~pc~cm$^{-3}$ per year.

Bursts from \frb{} have been detected at 
frequencies spanning from $\sim$0.3--8 GHz \citep{2020ApJ...896L..41C,2018ApJ...863....2G}. 
The bursting
activity of \frb{} does not seem to follow a Poissonian process, but rather
goes through phases of bursting activity and quiescence which can be better
explained with a Weibull distribution \citep{2018MNRAS.475.5109O}.
This dichotomy in activity could also be explained by the recently 
discovered apparent periodicity of \frb{} of 161 days with an
active window of 54\% 
\citep{2020arXiv200303596R,2020arXiv200803461C},
also detected in the repeating FRB 180916.J1058+65
with a period of 16 days and a 31\% activity window
\citep{2020Natur.582..351C}.

\frb{} is the first repeating FRB to be unambiguously localised 
to a host galaxy \citep{2017Natur.541...58C}, which is
a low-metallicity dwarf galaxy at a redshift 
of $z=0.193$ \citep{2017ApJ...834L...7T} with a stellar mass 
of $M_* \sim 1.3\times10^8$ M$_\odot$ and a star formation rate of 0.23 M$_\odot$
per year \citep{2017ApJ...843L...8B}. 
\frb{} is also coincident with a compact persistent radio source 
whose 
projected offset 
is $<$ 40 pc \citep{2017ApJ...834L...8M}.

The properties of \frb{} and its persistent radio source
have motivated a number of FRB models. 
Among the leading scenarios, FRBs are generated by flaring
magnetars within supernova remnants (SNRs).
Here, the magnetar flares collide with the surrounding medium,
producing shocks creating synchrotron maser emission,
resulting in FRB generation.
The main difference between these models lies in the nature
of the shocked material, being dominated by either
the magnetar wind nebula \citep[e.g.][]{2014MNRAS.442L...9L}, 
or by previous magnetar flares \citep[e.g.][]{2017ApJ...843L..26B,2019arXiv190807743B,2018ApJ...868L...4M}.

In this work we have observed \frb{} with the 
305-m William E. Gordon Telescope at the Arecibo Observatory (AO)
in Puerto Rico, USA, 
the Effelsberg 100-m Radio Telescope 
in Effelsberg, Germany, 
and the Karl G. Jansky Very Large Array (VLA) 
in New Mexico, USA,
to obtain RMs from its bursts in order to investigate its long-term RM evolution. 
In \S\ref{sect:obs} we describe our observations, data acquisition and search analysis.
In \S\ref{sect:res} we report sixteen new RM measurements of \frb{}, 
a long-term average \frb{} burst rate from our Effelsberg observations,
and discuss the properties of the detected bursts.
\S\ref{sect:impli} is dedicated to comparing our results to the theoretical
prediction of the RM evolution of an SNR 
from the works of \citet{2018ApJ...861..150P} and \citet{2018ApJ...868L...4M}, as
well as the Galactic center (GC) magnetar, \magn{} \citep{2018ApJ...852L..12D},
and in \S\ref{sect:disc} we interpret those results.
Finally, in \S\ref{sect:conc} we summarise our findings.

\section{Observations}
\label{sect:obs}


The observational setup and the data processing of each
telescope used in this work is detailed in their respective
subsections below.

We anticipated extremely high RM values from \frb{} bursts, 
and have thus observed at frequencies higher than the 
1.4-GHz band in order to avoid intra-channel depolarisation.

\subsection{Arecibo} \label{sect:obs_AO}
Data from the 305-m William E. Gordon Telescope at the 
Arecibo Observatory were acquired by using the C-band 
receiver at an observing frequency between $4.1$ and $4.9$\,GHz, 
composed of two orthogonal linear-polarization feeds.
The Puerto Rican Ultimate Pulsar Processing Instrument (PUPPI) 
backend recorded dual-polarisation data every $10.24$\,$\mu$s 
in $512$ frequency channels, each coherently dedispersed 
to $\text{DM}=557$\,pc\,cm$^{-3}$ to reduce intra-channel 
dispersive smearing to $< 2$\,$\mu$s.
The time and frequency resolution were reduced to $81.92$\,$\mu$s 
and $12.5$\,MHz, respectively, before searching for bursts.
We used \texttt{PRESTO}\footnote{\url{github.com/scottransom/presto}} 
\citep{2011ascl.soft07017R}
to create $200$ dedispersed time-series 
between $461$ and $661$\,pc\,cm$^{-3}$, which were searched by 
\texttt{single\_pulse\_search.py} with box-car filters ranging 
from $81.92$~$\mu$s to $24.576$~ms.
A large fraction of detections due to noise and radio frequency interference (RFI) were excluded 
by using dedicated software\footnote{\url{http://ascl.net/1806.013} 
\citep{2018ascl.soft06013M}} \citep{2018MNRAS.480.3457M}.
A `waterfall' plot of signal intensity as a function of time and 
frequency was produced and visually inspected for the rest of the detections.
The \texttt{DSPSR} package\footnote{\url{http://dspsr.sourceforge.net/}} 
\citep{2011PASA...28....1V} was used to create 
\texttt{PSRCHIVE}\footnote{\url{http://psrchive.sourceforge.net/}}  
\citep{2004PASA...21..302H}
files containing the full resolution data recorded by PUPPI. 

For each observation, 
the differential gain of the two receiver feeds was corrected for by analyzing with PSRCHIVE utilities a one-minute noise injection performed with a diode.
Additional calibrations, such as observations of a calibration source, were not performed and residual artifacts in the measured polarization fraction may be present.
RM values were calculated by using a technique called RM synthesis \citep{1966MNRAS.133...67B,2005A&A...441.1217B}, which reconstructs the Faraday dispersion function (FDF) of a source through a Fourier transform, in the implementation included in the \texttt{RM-tools} package.\footnote{\url{https://github.com/CIRADA-Tools/RM-Tools}}
We cleaned the resulting FDF by using a deconvolution algorithm \citep{2009IAUS..259..591H}.
RM uncertainties are estimated from the width of the peak in the FDF, rescaled by the maximum S/N.
The FDF of bursts detected on MJDs 58222 and 58712 (bursts 8, 19 and 20) shows signs of a poor polarisation calibration, namely symmetric RM peaks around the origin. 
This is likely caused by a delay calibration issue and,
while the RM measurements are still valid, the resulting linear polarisation fractions should be considered lower limits.
Conversely, the lack of an RM$= 0$\,rad~m$^{-2}$ peak for the bursts with $\sim 100$\% linear polarisation fractions indicates that cross-coupling between the X and Y polarisations does not have a significant effect on our results.  For these bursts, we also find no evidence for significant circular polarisation varying with radio frequency, which can occur if there is cross-coupling and poor calibration.
PA curves were calculated by de-rotating the data with \texttt{PSRCHIVE} at the RM value obtained for each burst.

\subsection{Effelsberg}

We have used the Effelsberg 100-m radio telescope to observe
\frb{} at 4--8~GHz
using the S45mm receiver
with a roughly two-week cadence for 2--3~hours each session from late 
2017 to early 2020,
totaling 115 hours.
The S45mm receiver has dual linear polarisation feeds.

The data were recorded with full Stokes information using two ROACH2 backends with each one
capturing 2~GHz of the band. The channel bandwidth is 0.976562~MHz across 4096
channels, with a 131~$\mu$s sampling rate. The recorded data were in a 
Distributed Acquisition and Data Analysis (DADA) 
format\footnote{\url{http://psrdada.sourceforge.net}}.
Before processing, Stokes \textit{I} was extracted from the data into a
SIGPROC filterbank\footnote{\url{http://sigproc.sourceforge.net}} 
format in order to perform the
initial burst searching.

Observations on 22nd October 2018 encountered a receiver issue, 
forcing us to use the S60~mm receiver instead.
The S60~mm receiver has an SEFD of 18~Jy, 500~MHz of bandwidth from 4.6 to 5.1~GHz,
0.976562~MHz channel bandwidth across 512 channels, and an 82~$\mu$s 
sampling rate.
The data were recorded as SIGPROC filterbanks. 

The data were searched for single pulses using the 
\texttt{PRESTO} 
software package. 
We used \texttt{rfifind} to identify RFI
in the data over two-second intervals
and to make an RFI mask which was applied to the data during searching. 
We used \texttt{PRESTO} to create dedispersed time-series of the data
from 0--1000~pc~cm$^{-3}$ in steps of 2~pc~cm$^{-3}$,
which were searched 
for single pulses
using \texttt{single\_pulse\_search.py} 
to convolve the time-series with
boxcar filters of varying widths to optimise the signal-to-noise of a burst. 
A pre-determined list of boxcar widths from \texttt{PRESTO} was used, where the widths
are multiples of the data sampling time. We searched for burst widths up to 19.6~ms
and applied a signal-to-noise threshold of 7.
DM-time and frequency-time plots of candidates were 
visually inspected to search for bursts.

For further RFI mitigation we calculated the modulation index of candidates. 
The modulation index assesses a candidate's fractional variations across the
frequency channels in order to discriminate between narrowband RFI and an
actual broadband signal \citep{2012ApJ...748...73S}. 
We applied this thresholding following \citet{2020MNRAS.493.5170H}.

If a burst was detected, we performed polarisation calibration
in order to obtain the RM, polarisation angle (PA), and degree of
polarisation of the burst. We used the 
\texttt{psrfits\_utils} package\footnote{\url{github.com/demorest/psrfits\_utils}}
to create a 
\texttt{psrfits}\footnote{\url{atnf.csiro.au/research/pulsar/psrfits\_definition/Psrfits.html}}
file containing the burst and used 
\texttt{PSRCHIVE} 
to calibrate the data by first dedispersing the burst data
using \texttt{pam}, then \texttt{pac} to polarisation calibrate those data
with noise diode observations. 
RM was obtained using RM synthesis, described in \S\ref{sect:obs_AO}.

\subsection{VLA}
FRB~121102 was observed with the VLA as part of a monitoring 
project (VLA/17B-283) from 2017~November to 2018~January.
Ten 1--hr observations were conducted at 2--4~GHz using 
the phased-array pulsar mode.
The VLA has dual circular polarisation feeds.
Data were recorded with full Stokes information with $8096\times 0.25$~MHz 
channels and $1024~{\rm \mu s}$ time samples.  
Each observation had $\approx 30$~min on-source.  Data were 
dedispersed at 150 trial DMs from $400-700~{\rm pc~cm}^{-3}$ 
and the resulting time-series were searched for pulses using 
the \texttt{PRESTO} \texttt{single\_pulse\_search.py}.

Polarisation calibration 
was done using the 10--Hz injected noise calibrator 
signal.  After polarisation calibration, 
the RMs 
were obtained using RM synthesis, as described in \S\ref{sect:obs_AO}.

\section{Observational Results}
\label{sect:res}

From our observations we have sixteen new RM measurements from \frb{} bursts:
1 from Effelsberg, 2 from the VLA, and 13 from Arecibo.
The details of our detections, along with previously reported RM values, 
are listed in Table \ref{tab:bursts}. 
The previously reported RM values from Arecibo \citep{2018Natur.553..182M}
and the GBT \citep{2018ApJ...863....2G} listed in Table \ref{tab:bursts}
are a global fit to multiple bursts from the same epoch.
Each burst is also assigned a numerical value for clarity.
The burst DMs in Table~\ref{tab:bursts} are obtained from
L-band bursts detected at Arecibo by linearly interpolating
their four-week averaged DM values, producing a mode
error value of 0.14~pc~cm$^{-3}$ for the interpolated values 
(Seymour et al., in prep.).
The L-band burst DMs were determined by maximising the
structure of the bursts and their 
sub-component alignment\footnote{\url{http://ascl.net/1910.004}
\citep{2019ascl.soft10004S}}.

\begin{table*}
    \centering
    \caption{Burst detections of \frb{} with measured RMs in chronological order. 
            \textit{From left to right:}  
            Burst number,
            barycentric burst arrival time in MJD
            (referenced to infinite frequency),
            width ($w$, full-width at half-maximum), 
            flux density ($S$), fluence ($F$), observed RM, DM,
            observing frequency, and telescope used.
            The burst DMs are approximated by linearly interpolating the four-week averaged DM values
            of the L-band bursts detected at Arecibo. This produces a mode error value of 0.14 pc cm$^{-3}$ for the interpolated values. The initial L-band DM values are determined by maximising their burst and sub-component alignment.
            Sub-bursts of multi-component bursts are further labeled 
            chronologically with lower-case letters.
            Previously reported bursts and bursts introduced in this work
            are separated by a horizontal line. 
            Abbreviations are AO: Arecibo Observatory,
                Eff: Effelsberg,
                GBT: Green Bank Telescope,
                VLA: Very Large Array.
            }
    \label{tab:bursts}
    \begin{tabular}{c c c c c c c c c }
    Burst & MJD & $w$  & $S$  & $F$     & RM$_\mathrm{obs}$            & DM            & Freq.  & Telescope \\
          &     & (ms) & (Jy) & (Jy~ms) & (rad~m$^{-2}$)& (pc~cm$^{-3}$)&(GHz) & \\
    \hline
    1 & 57747.12956--57747.17597  &&&& $102708(4)$    &  & 4.1--4.9  & AO $^\text{a,c}$ \\
    2 & 57748.12564--57748.17570  &&&& $102521(4)$    &  & 4.1--4.9  & AO $^\text{a,c}$\\
    3 & 57772.12903030  &&&& $103039(4)$    &  & 4.1--4.9  & AO $^\text{a}$\\
    4 & 57991.58013--57991.58330  &&&& $93573(24)$    &  & 4--8  & GBT $^\text{a,b,c}$\\
    \hline
    5 & 58069.31853200  & $4.49\pm0.09$ & $0.38\pm0.06$ & 1.69 & $86678(3)$   & $560.5$   & 2--4  & VLA \\
    6a & \multirow{3}{*}{58075.20058018}  & $1.65\pm0.07$ & $0.16\pm0.02$ & 0.26 & 
        \multirow{3}{*}{$86518(3)$} & \multirow{3}{*}{$560.6$} 
        & \multirow{3}{*}{2--4}  & \multirow{3}{*}{VLA} \\ 
    6b  &              & $1.95\pm0.08$ & $0.32\pm0.05$ & 0.63 &                   &           &       &\\
    6c  &              & $3.62\pm0.08$ & $0.56\pm0.08$ & 2.03 &                   &           &       &\\
    7 & 58215.86332798  & $0.34\pm0.01$ & $0.19\pm0.03$ & 0.06 & $70841(38)$    & $561.5$    & 4.1--4.9 & AO \\
    8a & \multirow{2}{*}{58222.85751812}  
        & $0.59\pm0.01$ & $0.32\pm0.05$ & 0.19 & 
        \multirow{2}{*}{$72090(22)$} & \multirow{2}{*}{$561.5$} 
        & \multirow{2}{*}{4.1--4.9}  & \multirow{2}{*}{AO}\\
    8b  &              & $0.34\pm0.04$ & $0.05\pm0.01$ & 0.02 &                   &           &       &\\
    9 & 58227.83201090  & $0.76\pm0.04$ & $0.08\pm0.01$ & 0.06 & $72062(62)$   & $561.6$     & 4.1--4.9  & AO\\
    10& 58228.63801964  & $0.69\pm0.08$ & $0.35\pm0.05$ & 0.24 & $72248(21)$  & $561.6$   & 4--8  & Eff \\ 
    11& 58234.81180934  & $0.39\pm0.03$ & $0.09\pm0.01$ & 0.04 & $73514(75)$   & $561.6$     & 4.1--4.9  & AO\\
    12& 58234.81642918  & $0.35\pm0.01$ & $0.20\pm0.03$ & 0.07 & $73358(38)$   & $561.6$     & 4.1--4.9  & AO\\
    13& 58243.77965432  & $0.52\pm0.01$ & $1.8\pm0.3$ & 0.96 & $71525(3)$   & $561.7$     & 4.1--4.9  & AO\\
    14a& \multirow{2}{*}{58244.77641721}  
        & $0.92\pm0.03$ & $0.13\pm0.02$ & 0.12 
        & \multirow{2}{*}{$71158(33)$} & \multirow{2}{*}{$561.7$} 
        & \multirow{2}{*}{4.1--4.9}  & \multirow{2}{*}{AO}\\
    14b  &              & $0.69\pm0.03$ & $0.12\pm0.02$ & 0.08 &                   &           &       &\\
    15& 58247.81273381  & $0.54\pm0.02$ & $0.13\pm0.02$ & 0.07 & $68937(72)$   & $561.7$     & 4.1--4.9  & AO\\
    16& 58677.60475978  & $0.30\pm0.02$ & $0.13\pm0.02$ & 0.04 & $69375(43)$   & $563.3$     & 4.1--4.9  & AO\\
    17& 58684.58367814  & $0.47\pm0.03$ & $0.09\pm0.01$ & 0.04 & $69524(58)$   & $563.2$     & 4.1--4.9  & AO\\
    18& 58684.58990897  & $0.25\pm0.02$ & $0.09\pm0.01$ & 0.02 & $69408(92)$   & $563.2$     & 4.1--4.9  & AO\\
    19a& \multirow{2}{*}{58712.47972031}  
        & $0.90\pm0.03$ & $0.27\pm0.04$ & 0.24 
        & \multirow{2}{*}{$66949(11)$} & \multirow{2}{*}{$563.1$}   
        & \multirow{2}{*}{4.1--4.9} & \multirow{2}{*}{AO}\\
    19b  &              & $0.207\pm0.001$ & $1.9\pm0.3$ & 0.40 &                   &           &       &\\
    20& 58712.48531398  & $1.89\pm0.09$ & $0.06\pm0.01$ & 0.11 & $67033(87)$   & $563.1$   & 4.1--4.9   & AO\\
    \multicolumn{7}{l}{$^\text{a}$\footnotesize{Results presented in \citet{2018Natur.553..182M}.}}\\
    \multicolumn{7}{l}{$^\text{b}$\footnotesize{Results presented in \citet{2018ApJ...863....2G}.}}\\
    \multicolumn{7}{l}{$^\text{c}$\footnotesize{Global fit to multiple bursts.}}
    \end{tabular}
\end{table*}

\subsection{Long-term Burst Rate at C-band at Effelsberg}

Previous surveys of \frb{} at frequencies between 4--8~GHz 
reported rates based on fewer observed hours \citep{2018ApJ...863..150S}
and anomalously high burst rates \citep{2018ApJ...863....2G}.
\citet{2018ApJ...863..150S} detected three bursts from observing at 
4.6--5.1~GHz for 22 hours consisting of 10 observing epochs spanning 
five months using the Effelsberg telescope. 
\citet{2018ApJ...863....2G} detected 21 bursts in a single six-hour
observation, observing at 4--8~GHz at the Green Bank Telescope.  
Furthermore, \citet{2018ApJ...866..149Z} re-searched the data
from \citet{2018ApJ...863....2G} using a convolutional neural network
and detected an additional 72 bursts within the data.

Our Effelsberg survey spans over two years of observing \frb{}
for 2--6~hrs at a time 
at 4--8~GHz with a two-week cadence, amounting to 115 hours of observations. 
Included here are 10 hours of observations presented in \citet{2020MNRAS.496.4565C}.
We can therefore report a robust,
long-term average burst rate of \frb{} in this frequency range of 
$0.21^{+0.49}_{-0.18}$~bursts/day (1-sigma error)
above a fluence of $0.04 \; (w/\mathrm{ms})^{1/2}$~Jy~ms
for a burst width of $w$~ms.
We list the details of the surveys discussed here in Table~\ref{tab:surveys}.

A caveat to our observed burst rate is the suspected periodic activity
of \frb{} \citep{2020arXiv200303596R}.
Roughly $40\%$ of our Effelsberg observations were performed during
suspected inactivity of \frb{}, 
which if true would affect the observed burst rate. 
No bursts were detected during this state of inactivity.
Including only observations while \frb{} is active,
the average burst rate becomes $0.35^{+0.80}_{-0.29}$~bursts/day
above a fluence of $0.04 \; (w/\mathrm{ms})^{1/2}$~Jy~ms.

The observed burst rates of \frb{} also seem to be frequency dependent,
with the rate being lower at higher frequencies.
At 1.4~GHz the \frb{} burst rate has been observed to be
$8\pm3$~bursts/day above a fluence of 0.08~Jy~ms for 1~ms
burst widths \citep{2020arXiv200803461C}.

\begin{center}
\begin{table*}
    \begin{center}
    \caption{\frb{} surveys at frequencies between 4--8~GHz. 
            \textit{From left to right:}
            Survey, number of bursts, number of hours observed,
            number of observing epochs, frequency range, and telescope used.
            All surveys except \citet{2018ApJ...863..150S} recorded
            full Stokes data.
            Abbreviations are AO: Arecibo Observatory, Eff: Effelsberg, GBT: Green Bank Telescope.}
    \label{tab:surveys}
    \begin{tabular}{c c c c c c}
    Survey & No. bursts & No. hours & No. epochs & Freq.~(GHz) & Telescope \\
    \hline
    \citet{2018ApJ...863..150S} & 3 & 22 & 10 & 4.6--5.1 & Eff \\
    \citet{2018Natur.553..182M} & 16 & 13 & 12 & 4.1--4.9 & AO \\
    \citet{2018ApJ...863....2G} & 21 & 6 & 1 & 4--8 & GBT \\ 
    \citet{2018ApJ...866..149Z}$^\text{a}$ & 93 & 6 & 1 & 4--8 & GBT \\
    This work & 1 & 115 & 35 & 4--8 & Eff \\
    \multicolumn{6}{l}{\footnotesize{$^\text{a}$Re-searching of data from \citet{2018ApJ...863....2G}}}
    \end{tabular}
    \end{center}
\end{table*}
\end{center}

\subsection{Burst Properties}

\begin{figure*}
    \centering
        \includegraphics[width=.24\textwidth,trim=58 70 175 75,clip]{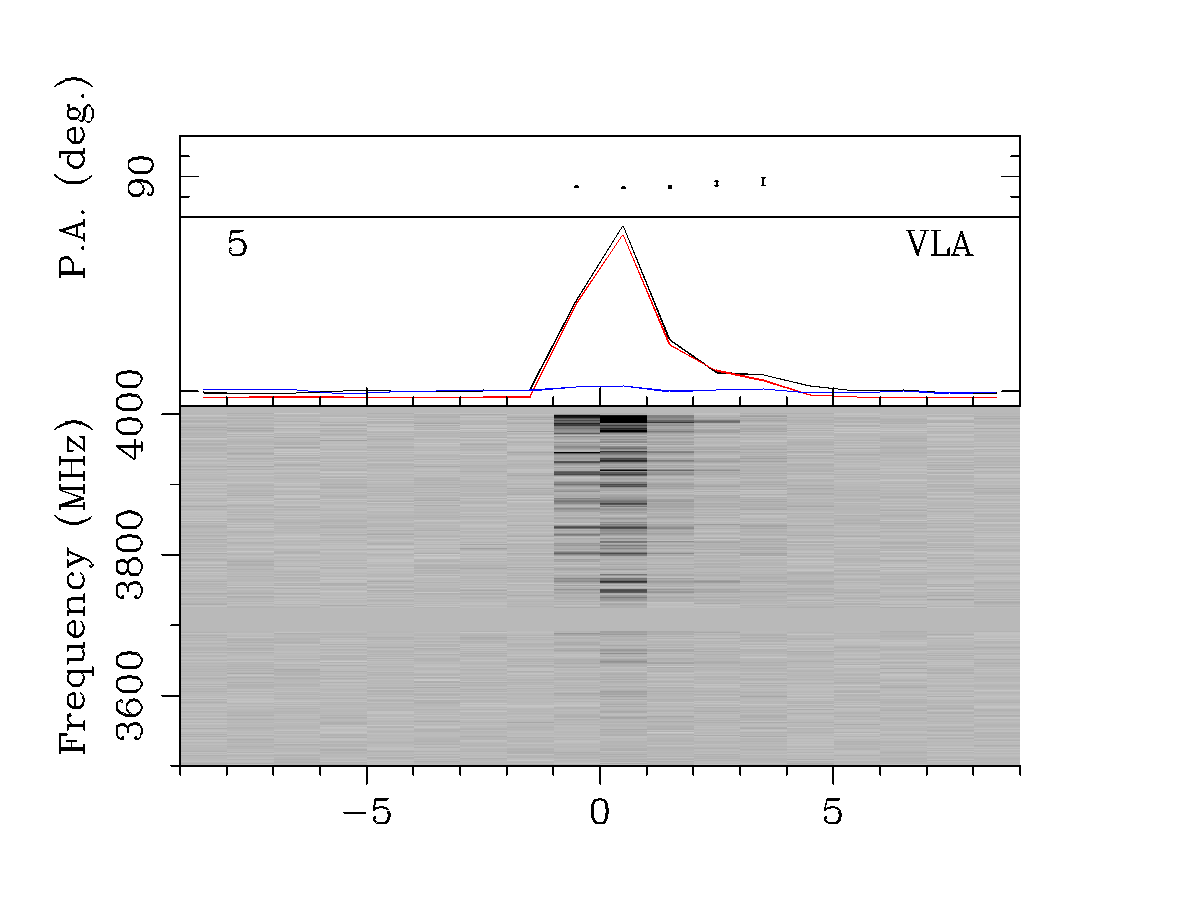}
        \includegraphics[width=.24\textwidth,trim=60 70 175 75,clip]{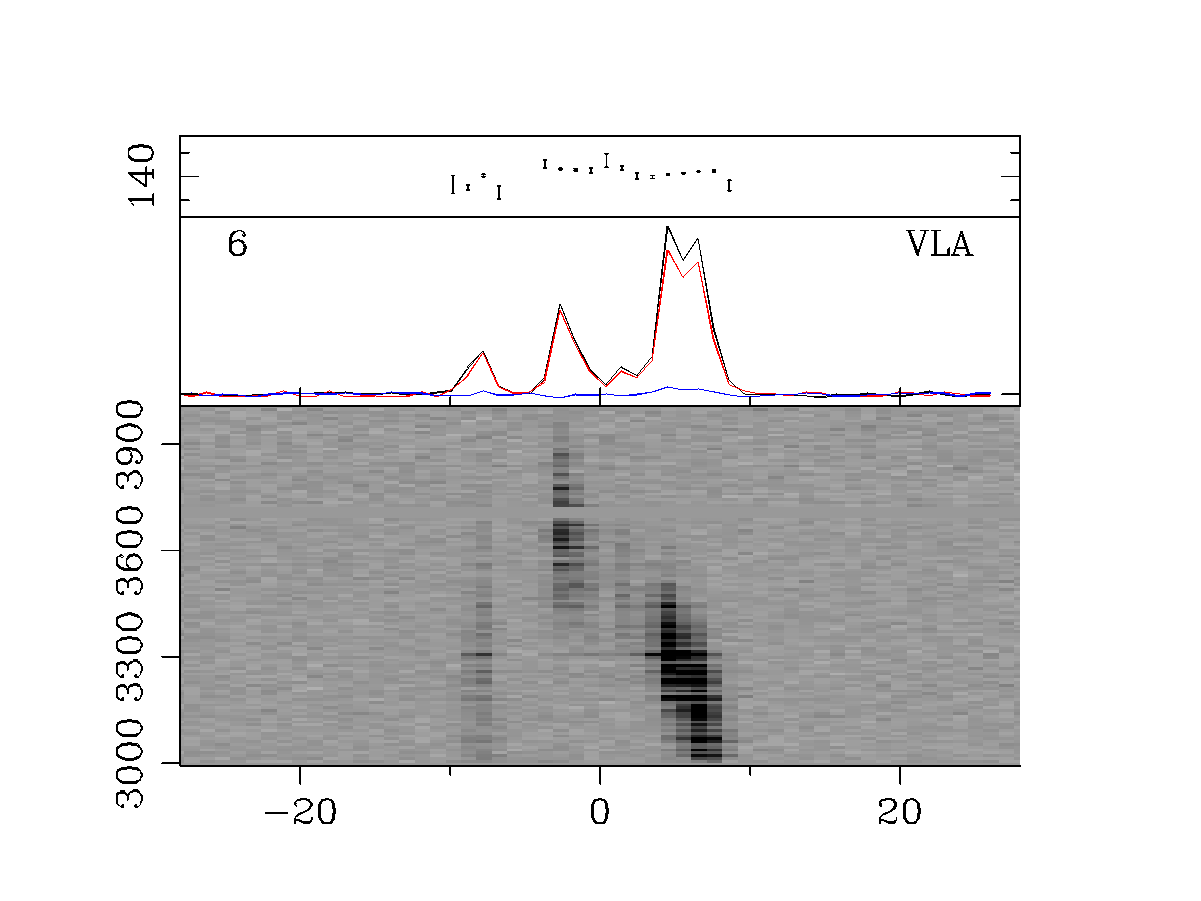}
        \includegraphics[width=.24\textwidth,trim=60 70 175 75,clip]{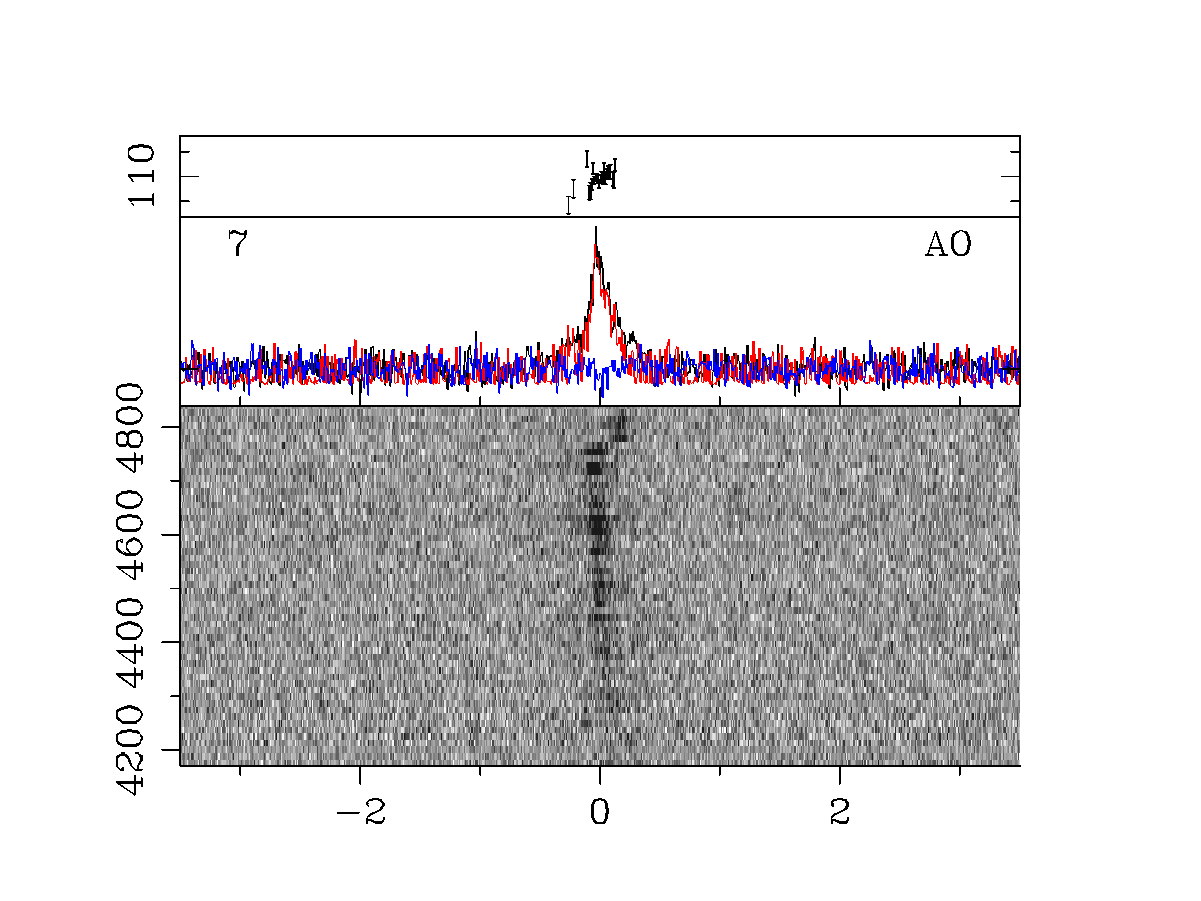}
        \includegraphics[width=.24\textwidth,trim=60 70 175 75,clip]{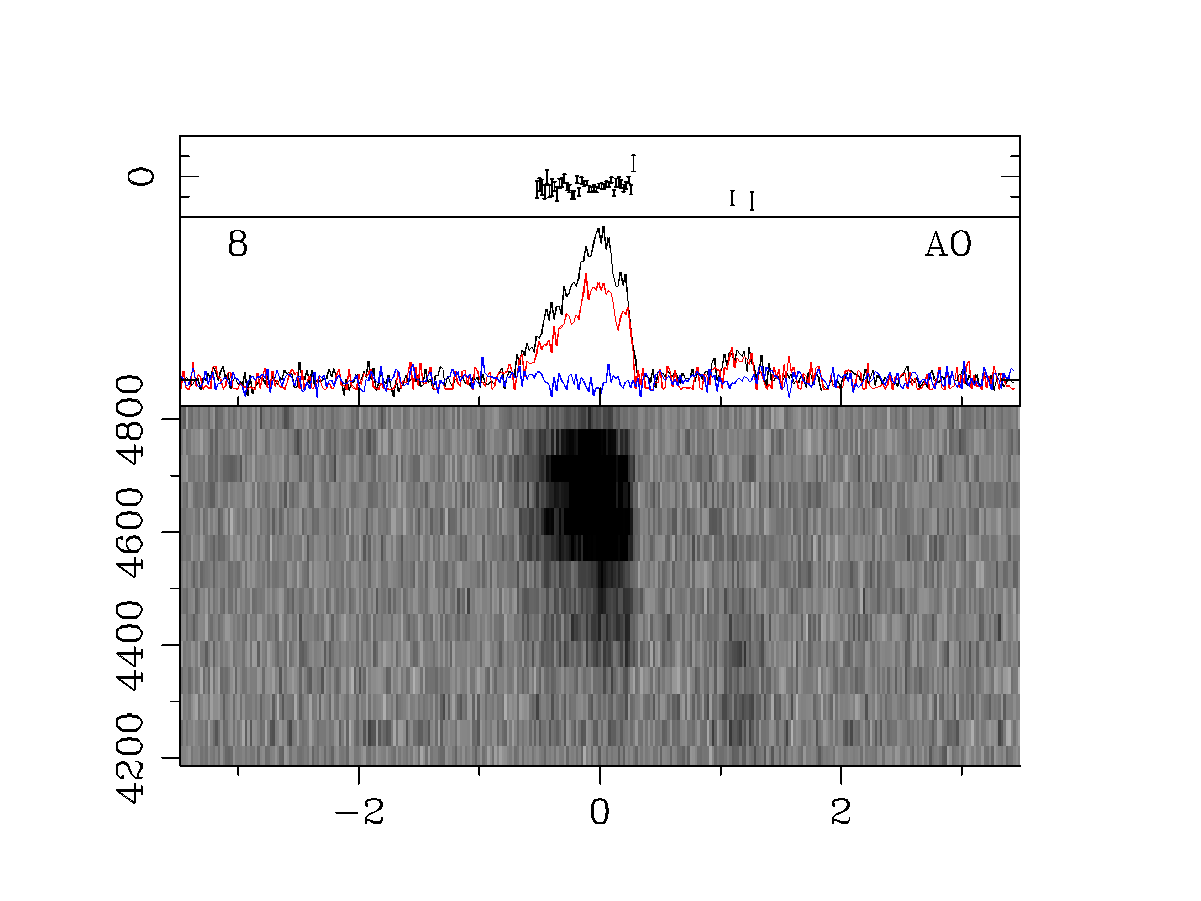}
        \includegraphics[width=.24\textwidth,trim=58 70 175 75,clip]{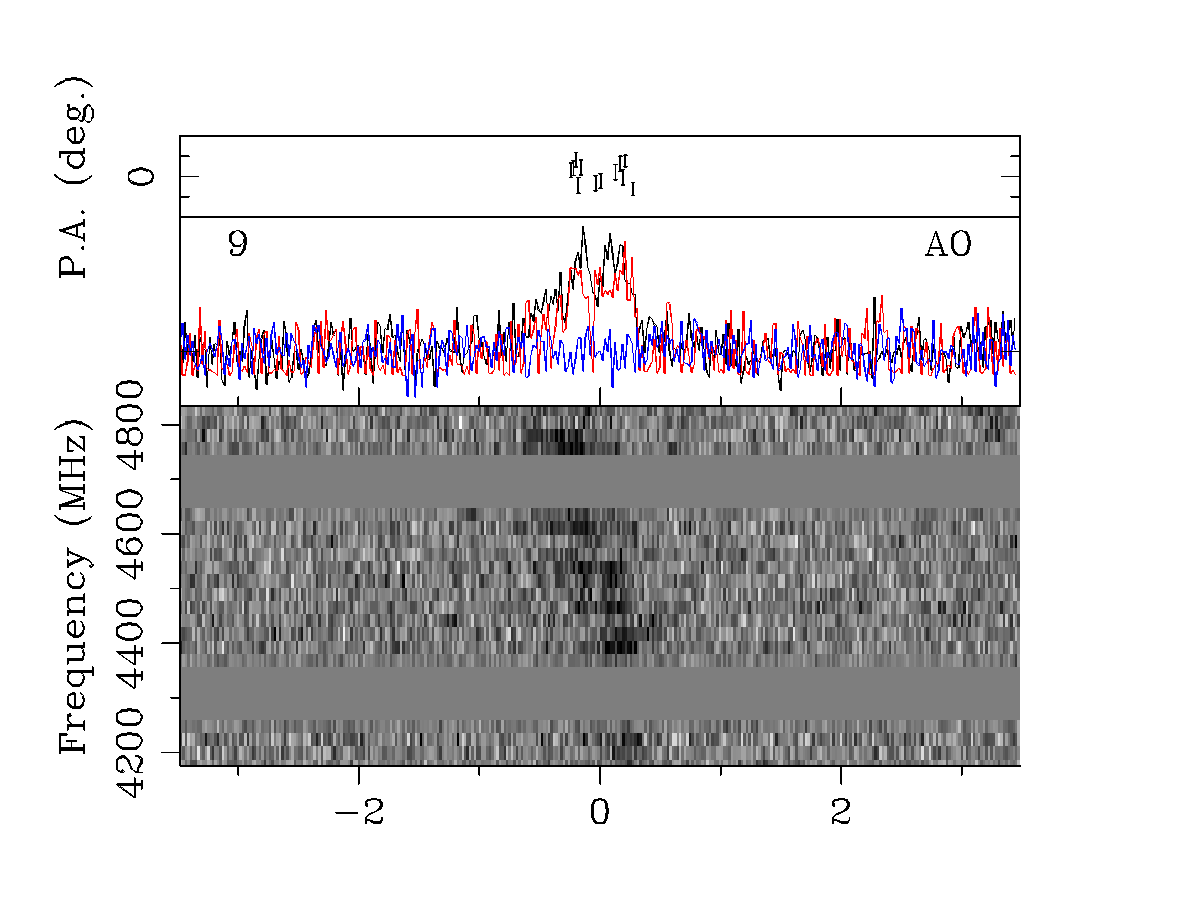}
        \includegraphics[width=.24\textwidth,trim=60 70 175 75,clip]{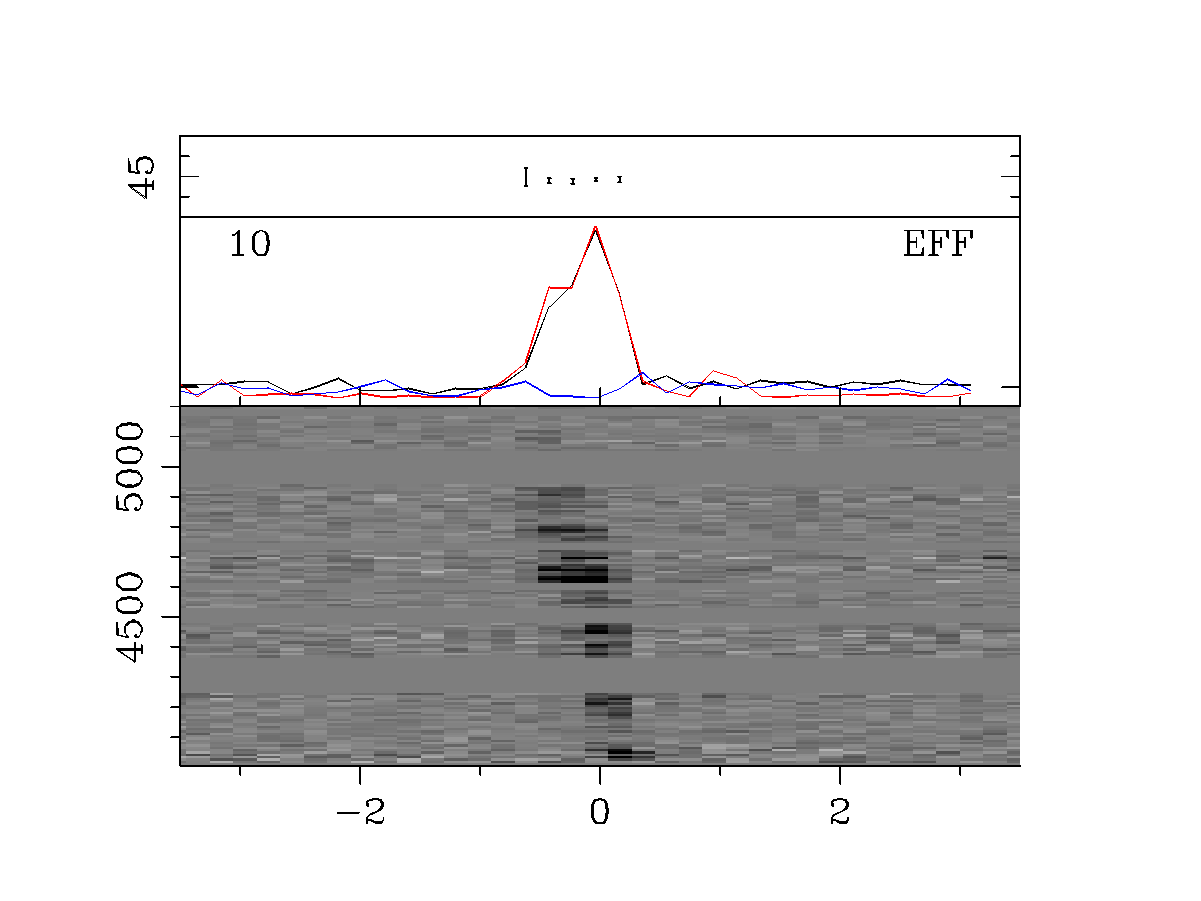}
        \includegraphics[width=.24\textwidth,trim=60 70 175 75,clip]{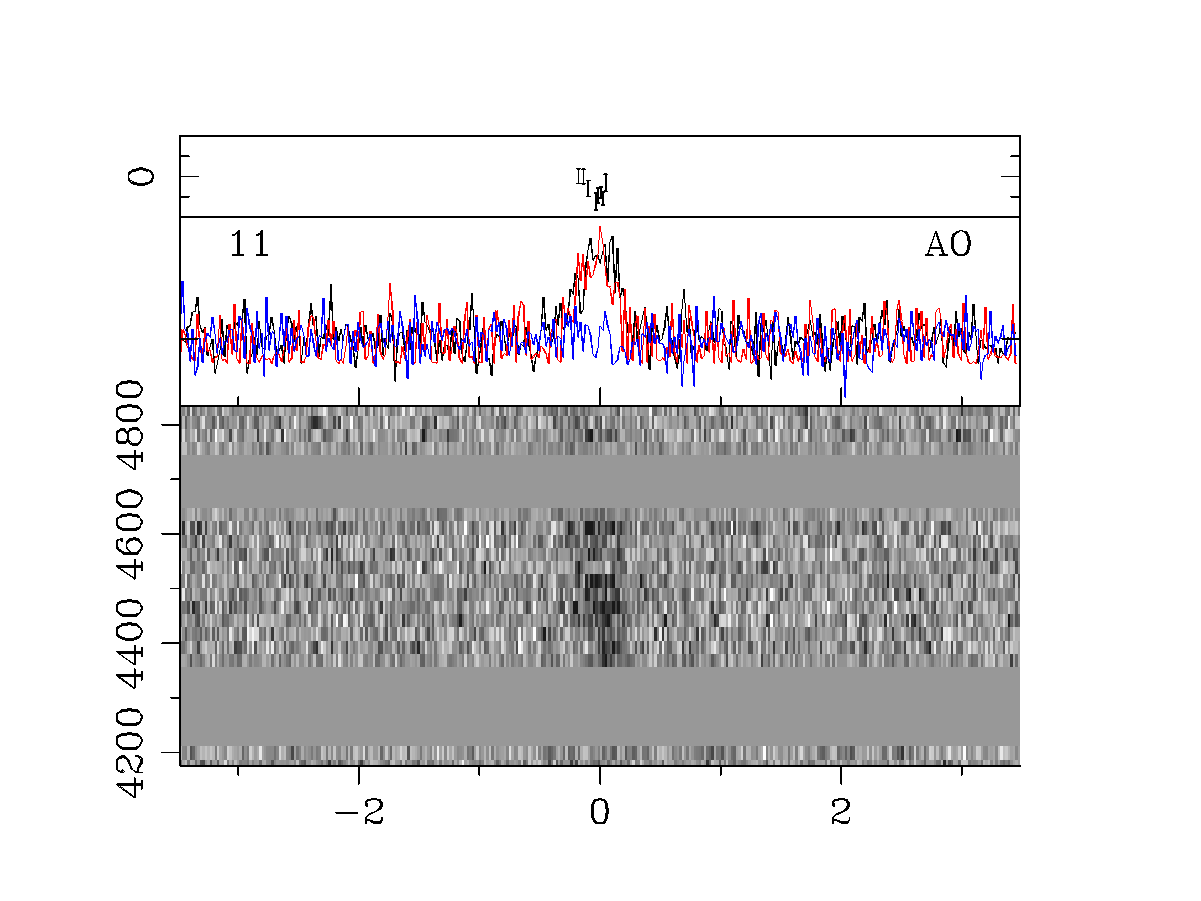}
        \includegraphics[width=.24\textwidth,trim=60 70 175 75,clip]{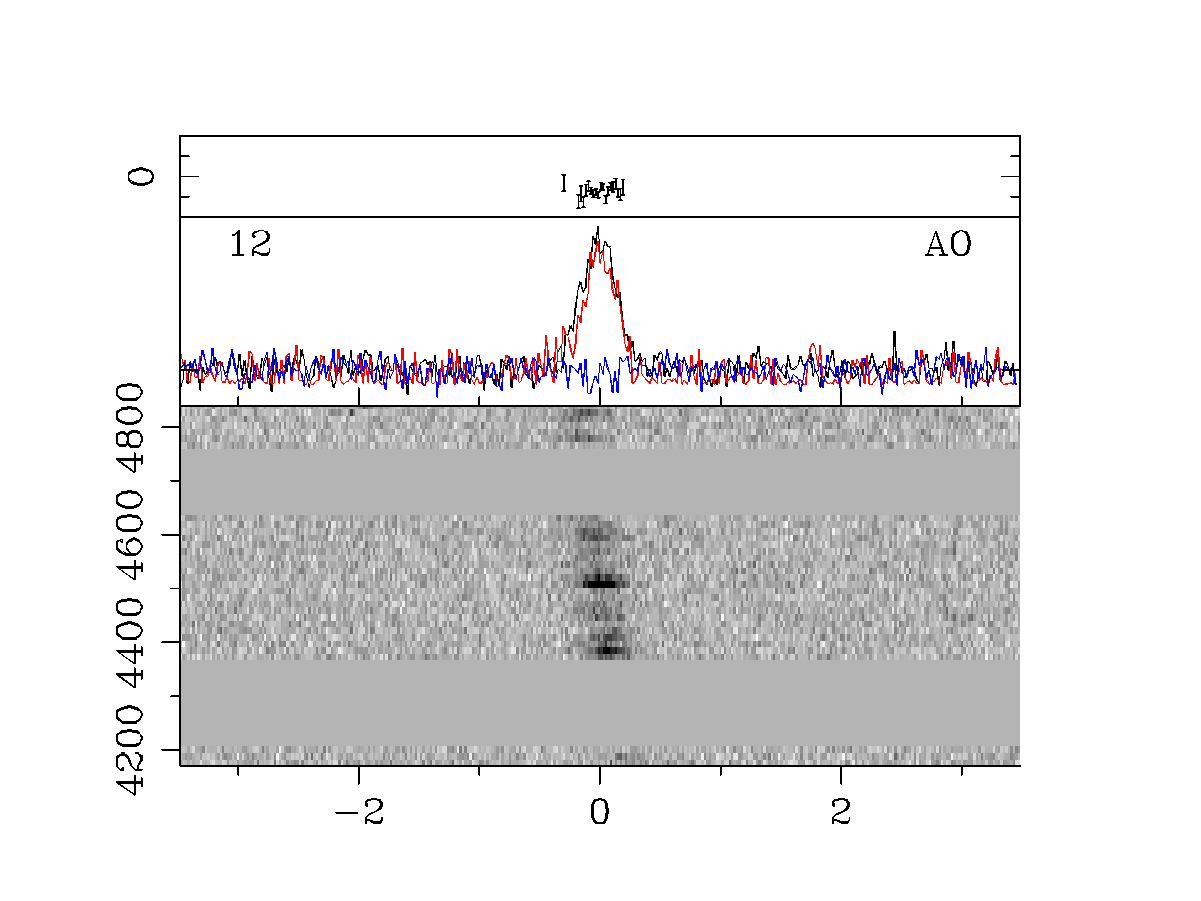}
        \includegraphics[width=.24\textwidth,trim=58 70 175 75,clip]{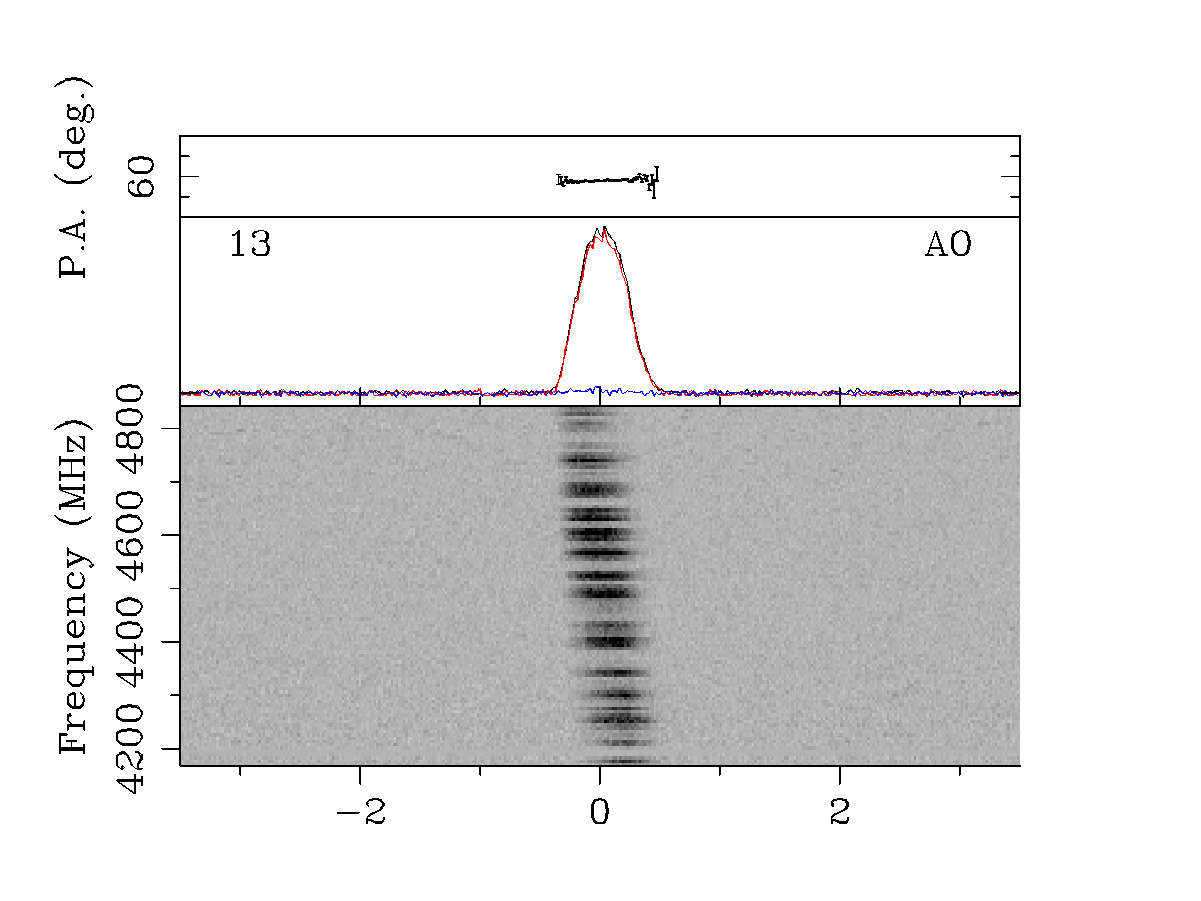}
        \includegraphics[width=.24\textwidth,trim=60 70 175 75,clip]{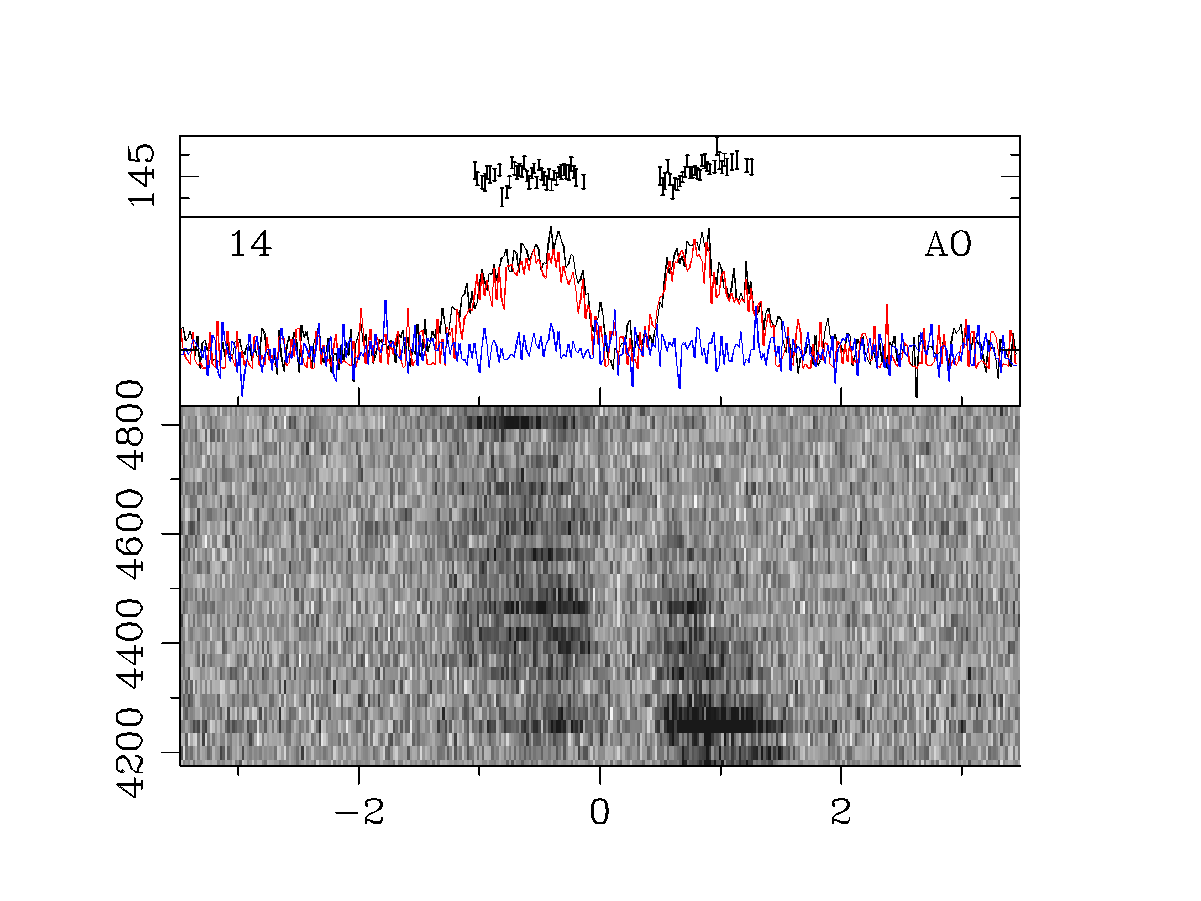}
        \includegraphics[width=.24\textwidth,trim=60 70 175 75,clip]{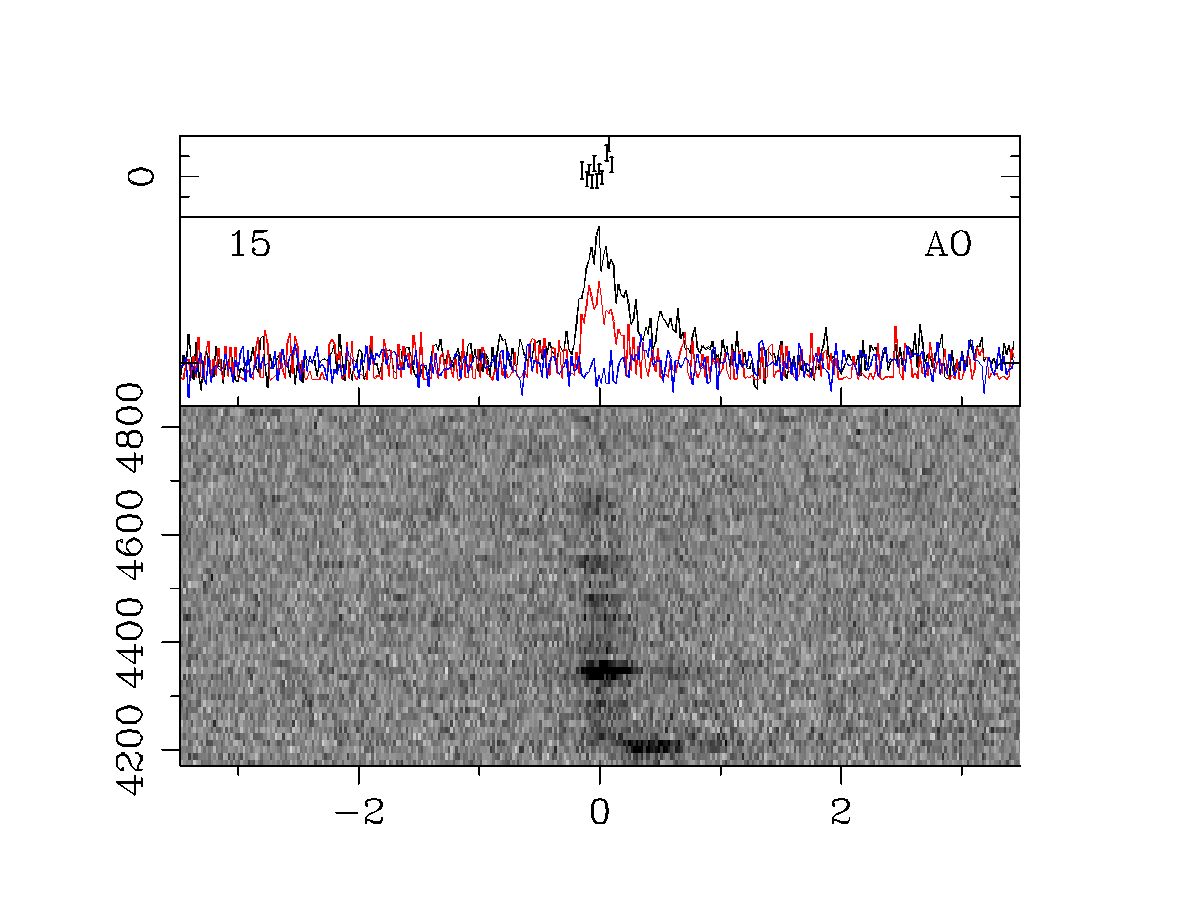}
        \includegraphics[width=.24\textwidth,trim=60 70 175 75,clip]{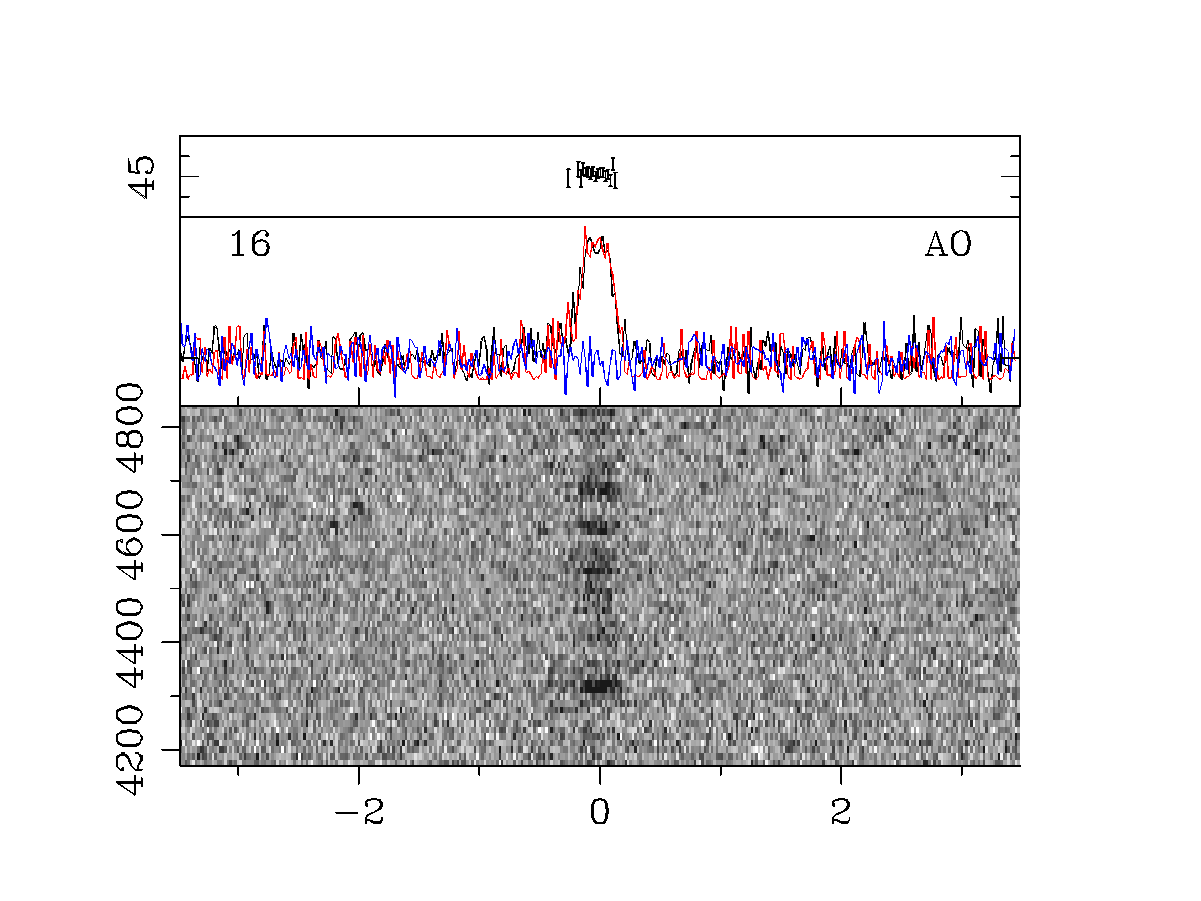}
        \includegraphics[width=.24\textwidth,trim=58 33 175 75,clip]{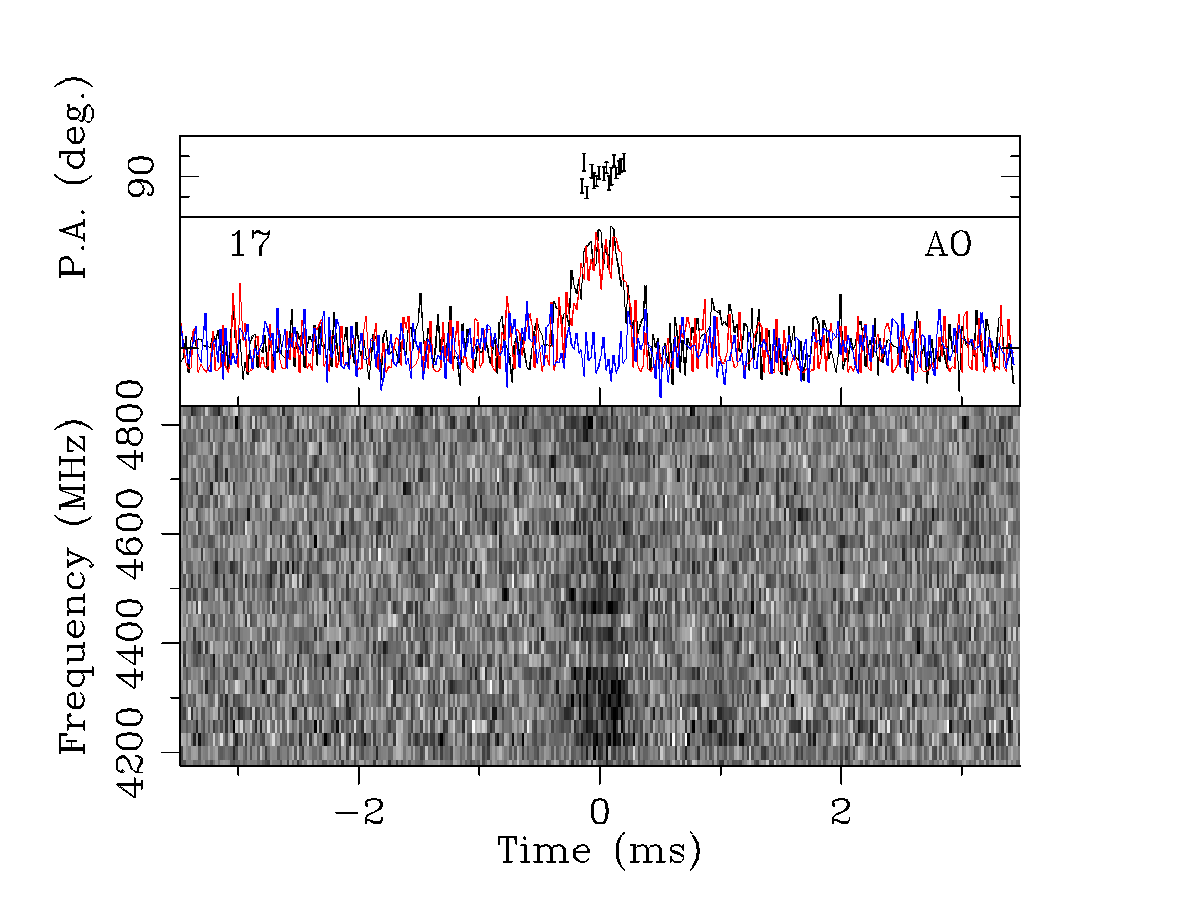}
        \includegraphics[width=.24\textwidth,trim=60 33 175 75,clip]{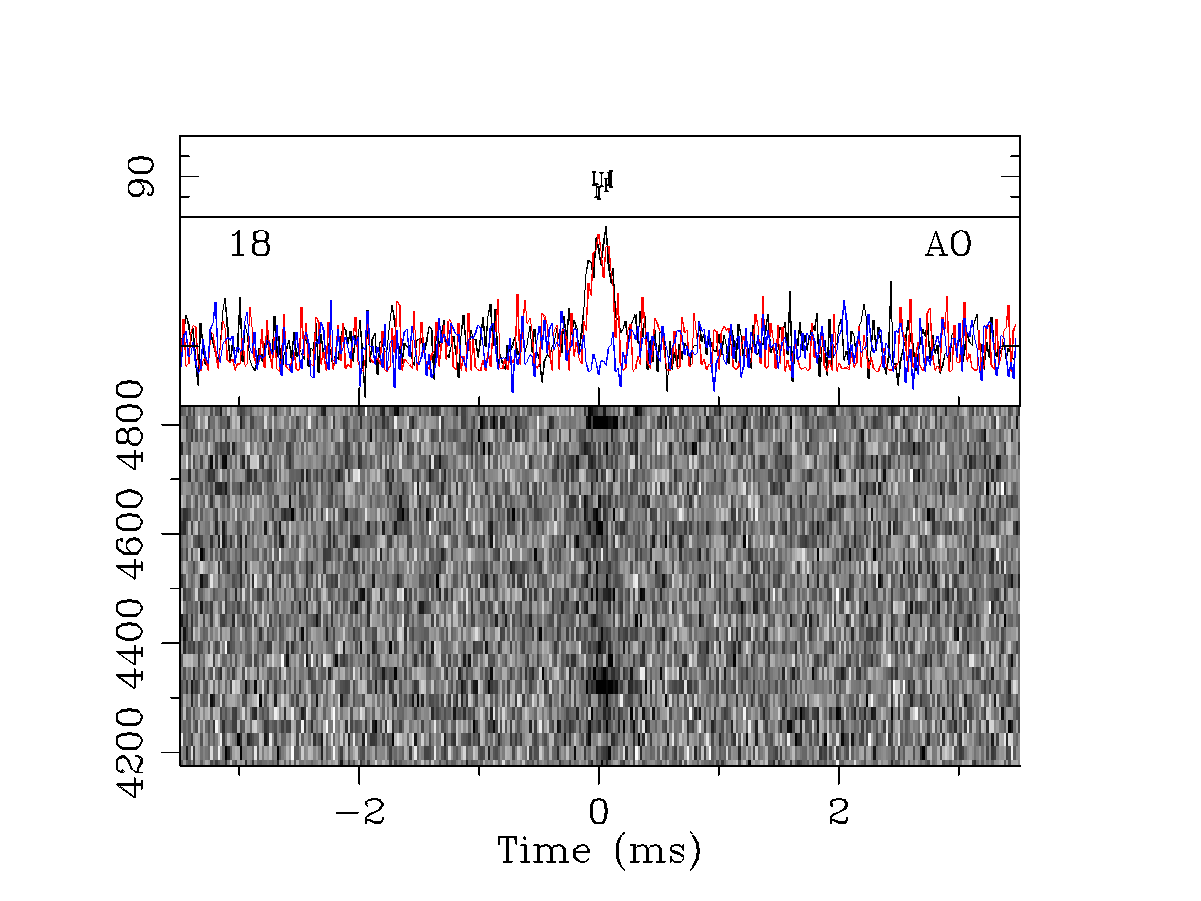}
        \includegraphics[width=.24\textwidth,trim=60 33 175 75,clip]{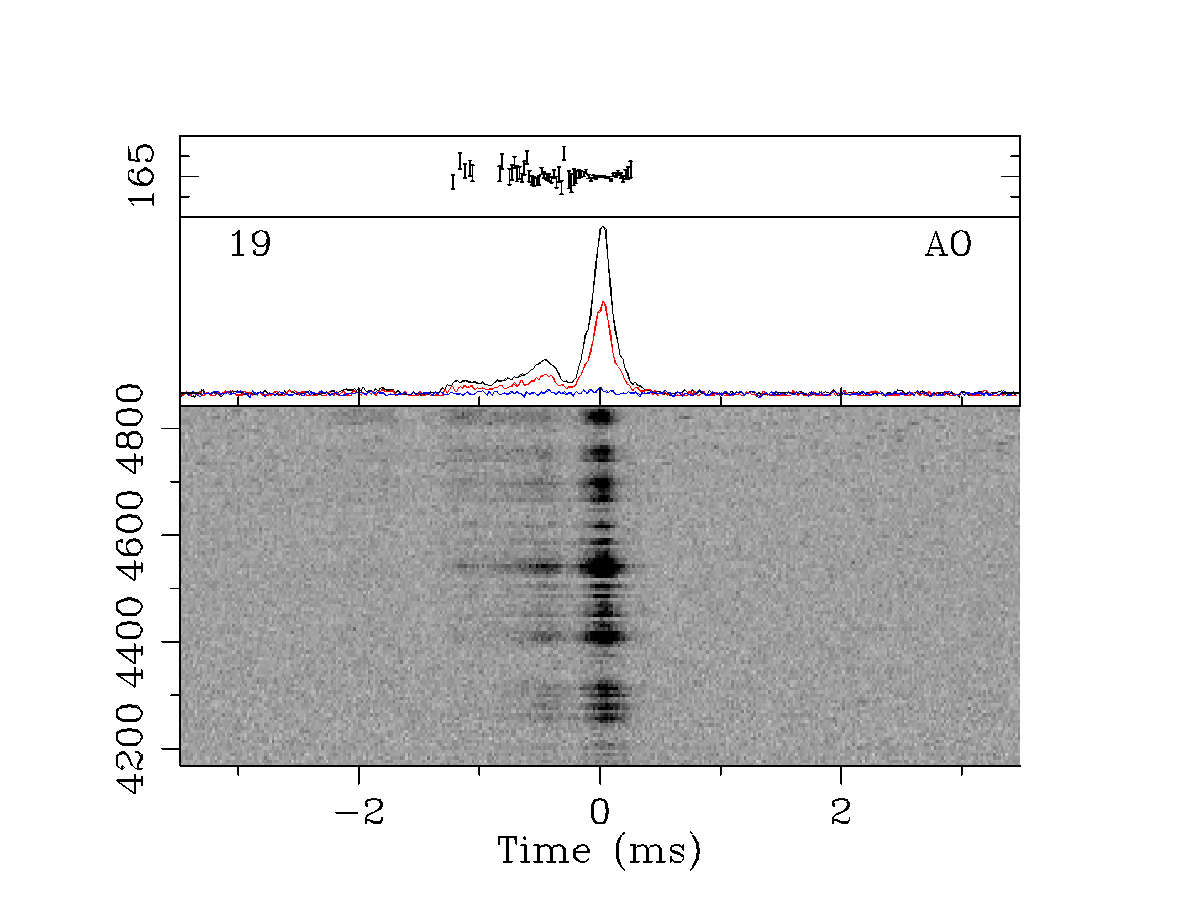}
        \includegraphics[width=.24\textwidth,trim=60 33 175 75,clip]{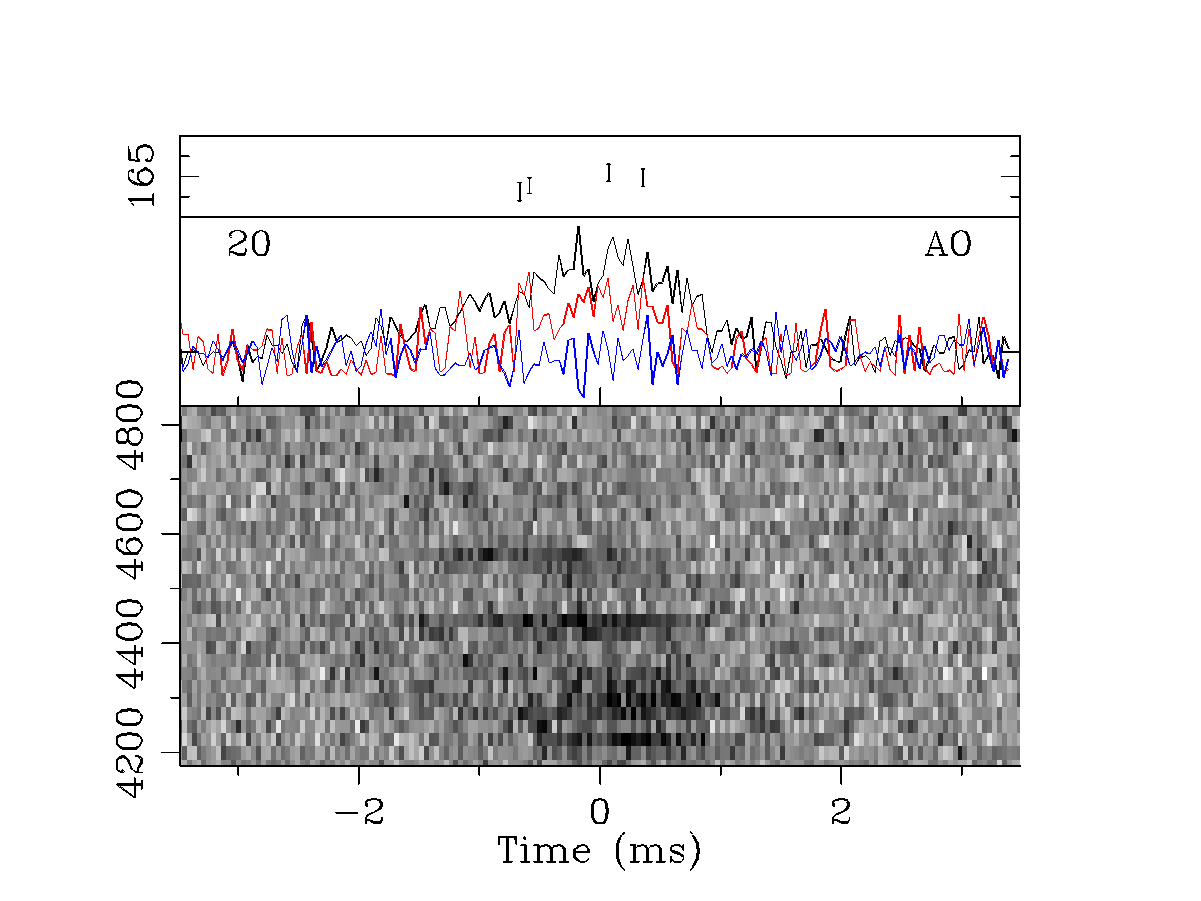}
    \caption{
            Dynamic spectra of the bursts detected with Arecibo,
            VLA, and Effelsberg in a chronological order,
            dedispersed to their respective DMs listed in Table \ref{tab:bursts}.
            On top of each spectrum is plotted the profile of the burst (in black),
            linear polarisation (red), and circular polarisation (blue), as well
            as the polarisation angle (PA). The PA range in each panel is 60 degrees.
            Each panel is labeled with the corresponding burst number from Table \ref{tab:bursts}
            and the telescope at which the burst was detected.
            Bursts 8, 15, 19, and 20 suffer from 
            delay calibration issues,
            resulting in unreliable polarisation fractions (see \S\ref{sect:obs_AO}).
            }
    \label{fig:burstfig}
\end{figure*}

We plot the dynamic spectra, polarisation profile, and polarisation angles (PAs)
of our detected bursts in Fig.~\ref{fig:burstfig}.
The PA is equal to 
$\mathrm{RM} \lambda^2 + \mathrm{PA}_\mathrm{ref}$,
where $\lambda$ is the observing wavelength,
and $\mathrm{PA}_\mathrm{ref}$ is a reference angle
at a specific frequency (central observing frequency
in our case).
The PAs are flat across each burst,
as has been seen previously from \frb{} 
\citep{2018Natur.553..182M,2018ApJ...863....2G}. 
We do not discuss PA changes over time, as we did not observe an 
absolute calibrator for polarisation.
In the absence of an absolute calibrator we
cannot compare PAs across multiple telescopes, and a more detailed discussion is outside the scope of this work.

The bursts are mostly $\sim$~$100\%$ linearly polarised. 
Bursts from \frb{} have been consistently $\sim100\%$ linearly
polarised since its first polarisation measurement in late 2016 
\citep{2018Natur.553..182M}, which suggests a stability in
its emission process.
The Arecibo bursts at MJD 58222, 58247, and 58712
(bursts 8, 15, 19, and 20)
are not fully linearly polarised, which is 
uncharacteristic for \frb{}, 
and can be attributed to polarisation calibration issues (see \S\ref{sect:obs_AO}).

The circular polarisation fraction is on average $4\%$ for
all the bursts and $17\%$ by summing the absolute Stokes \textit{V}
values. The Stokes \textit{V} on-pulse and off-pulse standard deviation
is the same for the majority of the bursts, indicating that
the circular polarisation is consistent with noise, 
therefore the circular polarisation fractions quoted above are upper limits.
For the VLA and Effelsberg bursts (number 5, 6, and 10)
the on and off-pulse statistics differ, 
where the on-pulse standard deviation is larger by a factor of two,
and we obtain circular polarisation fractions for Stokes \textit{V}
and absolute Stokes \textit{V} of $5\%$/$5\%$, $2\%$/$4\%$, and $3\%$/$11\%$
for bursts 5, 6, and 10, respectively.
The $100\%$ linear polarisation and the lack of circular polarisation 
for the majority of bursts
indicates that little to no Faraday rotation conversion, where 
linear polarisation is converted to circular in a magneto-ionic environment 
\citep{2019MNRAS.485L..78V,2019ApJ...876...74G}, occurs at our observing
frequencies.

The VLA burst on MJD 58075 (burst 6) exhibits a triple component profile.
The second and third components exhibit
a downward drift in frequency, a feature predominantly observed from 
repeating FRBs \citep[e.g.][]{2019ApJ...876L..23H,2020arXiv200110275T}.
An apparent upward drift in frequency between the first two components can be seen,
and the first component has a different PA than the other two.
While the temporal spacing between the components is not large,
the difference in PAs between the first component and the other 
two might suggeest that these are in fact two separate bursts.

The Effelsberg burst at MJD 58228 (burst 10)
was only detected between 4--5.2~GHz of the
4--8~GHz bandwidth. 
We were affected by strong edge effects in the bandpass,
resulting in an uneven frequency response across the bandwidth.
Thus we are uncertain of whether the burst frequency envelope
is inherent to the burst or due to the bandpass.

\subsection{Dispersion and Rotation Measures of \frb{}}

\begin{figure*}
    \centering
    \includegraphics[width=\textwidth]{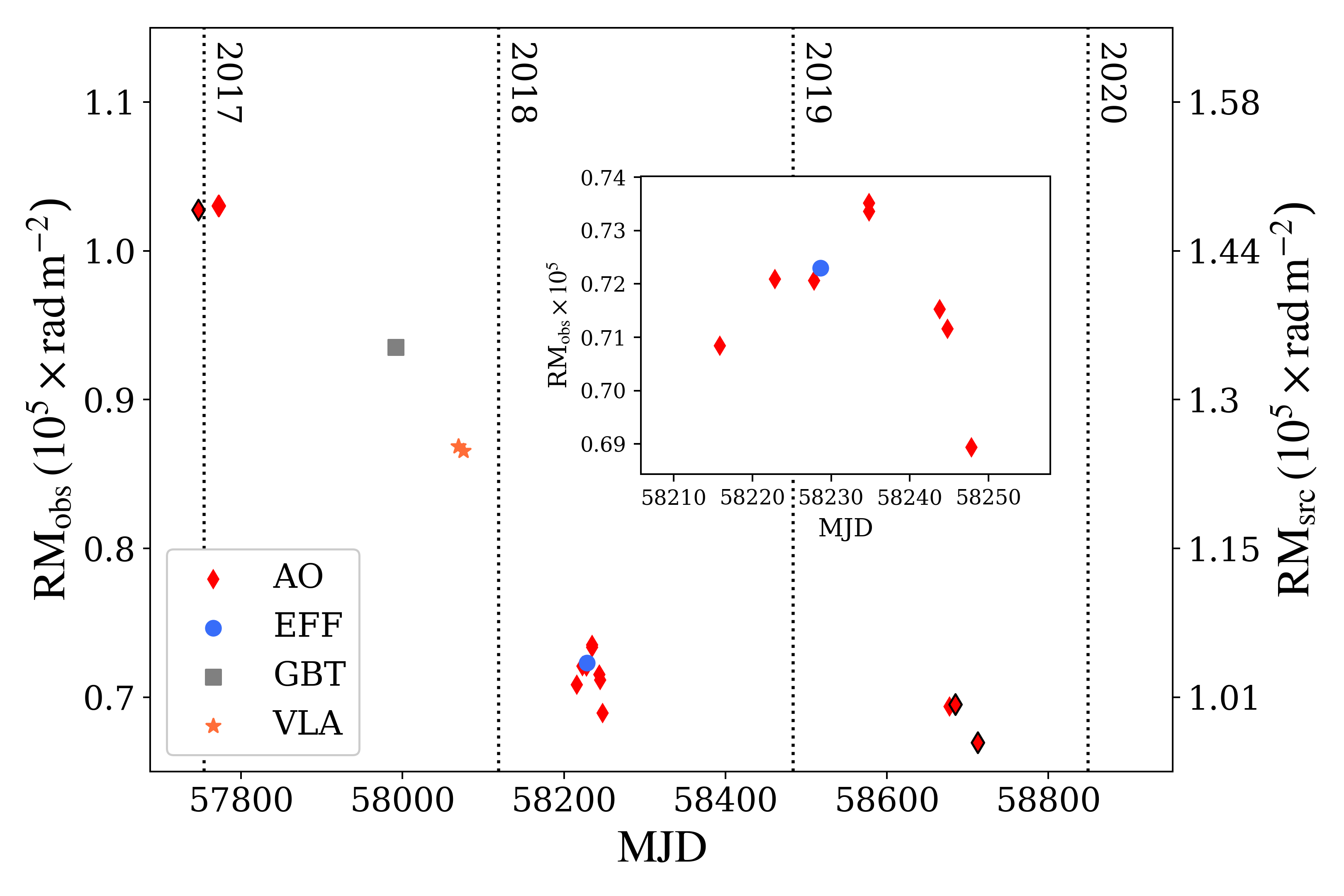}
    \caption{The 20 RMs of \frb{} as a function of time
            in MJD. 
            The left y--axis shows the observed RM
            and the right y--axis shows the RM in the
            source frame of \frb{}.
            Different markers indicate at which telescope
            the burst was detected. 
            Markers with a black edge indicate two bursts
            whose markers would otherwise overlap due to 
            being too close in time and RM.
            The horizontal dotted lines show the start of each
            calendar year.
            The inset gives a closer look at the cluster of bursts
            around MJD $\sim58230$ (when a high-cadence observing
            campaign was performed).
            The observed rotation measure uncertainties are not
            large enough to exceed the boundaries of the markers.
            Abbreviations are AO: Arecibo Observatory,
            Eff: Effelsberg,
            GBT: Green Bank Telescope,
            VLA: Very Large Array.
            The points near MJD 57800 and 58000 are data from
            \citet{2018Natur.553..182M} 
            and \citet{2018ApJ...863....2G} respectively.}
    \label{fig:RM_time}
\end{figure*}

\begin{figure}
    \centering
    \includegraphics[width=\columnwidth]{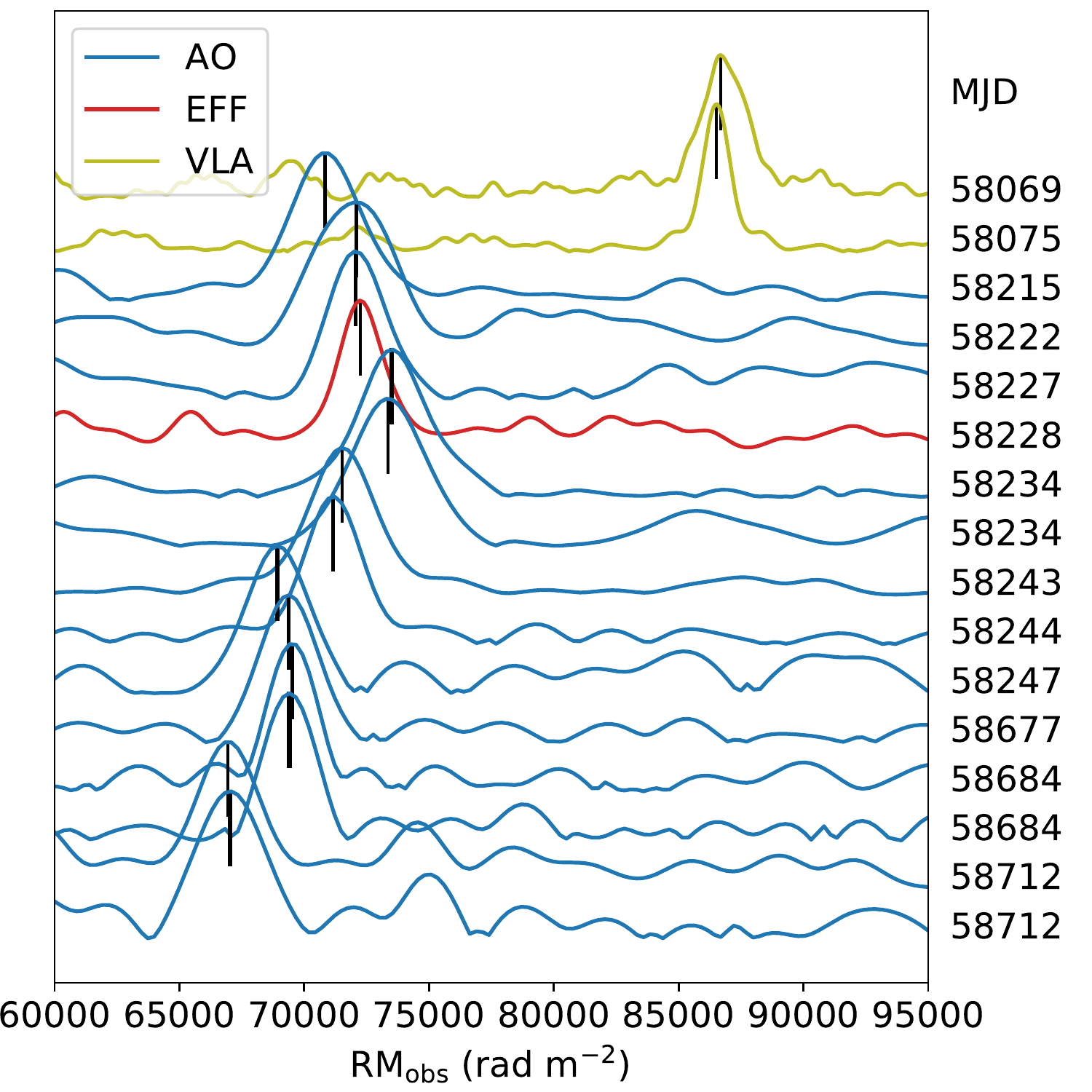}
    \caption{
            Faraday dispersion function of the bursts
            presented in this work, with different
            colors for different telescopes.
            Each curve is normalised to a unitary peak
            for clarity and the resulting RM value is
            indicated with a vertical black line whose
            thickness is the $1-\sigma$ uncertainty.
            The burst MJDs are reported on the right.}
    \label{fig:FDF}
\end{figure}

The RMs we obtained from our bursts are listed in Table~\ref{tab:bursts}.
We plot the RMs over time in Fig.~\ref{fig:RM_time}. 
The observed RM of \frb{} has dropped by 34\% over
2.6 years from $\sim 10^5$~rad~m$^{-2}$ to 
$\sim 6.7\times10^4$~rad~m$^{-2}$.
As Fig.~\ref{fig:RM_time} shows, the drop in RM has not
been steady over time. 
From MJD 57757 to MJD 58215 (bursts 1--7), the
RM decreased rapidly to $\sim7\times10^4$~rad~m$^{-2}$ 
and has declined only slightly ($\sim5000$~rad~m$^{-2}$) since then.
An FDF plot of the bursts reported here can be seen in Fig.~\ref{fig:FDF}.

Within a 32-day timespan the observed RM of \frb{}
exhibited significant short-timescale variations (bursts 7--15).
At epochs separated by a week, the RM increased by $\sim1000$~rad~m$^{-2}$
(bursts 7--8). 
For three epochs during the following week, 
the RM remained stable between bursts 8--10,
before increasing again by $\sim1000$~rad~m$^{-2}$ a week later (bursts 11--12).
During three epochs in the following two weeks,
the RM was observed to drop rapidly by a total of $\sim4500$~rad~m$^{-2}$ 
(bursts 12--15).
This short-timescale behaviour can be seen in the inset of Fig~\ref{fig:RM_time}.
No RM measurement is available between MJDs 58247 and 58677 (430 days),
but the RMs are consistent with each other at these dates (bursts 15--16). 
Another drop in RM of $\sim2000$~rad~m$^{-2}$ can be seen between bursts 18 and 19,
separated by 28 days.

Only minor changes in DM have been observed
during the observed RM evolution of \frb{}.
While the RM decreased significantly, the DM has
increased by $\sim4$~pc~cm$^{-3}$,
from $559.7\pm0.1$~pc~cm$^{-3}$ \citep{2018Natur.553..182M}
up to $563.3$~pc~cm$^{-3}$ from the aforementioned
linear interpolation of L-band burst DMs used in
this work.


\citet{2018Natur.553..182M} constrained the average magnetic field along
the LoS in the region which Faraday rotation occurs, 
$\langle B_\parallel \rangle$, between 0.6~mG and 2.4~mG using their measured \frb{}
RM in the source frame of 
$\mathrm{RM}_\mathrm{src} \sim1.4\times10^5$~rad~m$^{-2}$ and
the estimated host DM contribution of 
DM$_\mathrm{host}~70-270$~pc~cm$^{-3}$ \citep{2017ApJ...834L...7T}.
From a measured DM and RM, $\langle B_\parallel \rangle$ can be calculated,
ignoring sign reversals, as
\begin{equation}
\label{eq:Bfield}
    \langle B_\parallel \rangle = 1.23\; \mathrm{RM}_\mathrm{src}/
                   \mathrm{DM}_\mathrm{host} \mathrm{~}\mu\mathrm{G}.
\end{equation} 
The most recent DM and RM values in our sample yield 
$\langle B_\parallel \rangle = 0.4$--$1.6\mathrm{~mG}$.
This is a lower limit as the DM in the Faraday rotating
region could be much lower.

\section{Implications for source scenarios}
\label{sect:impli}

We explore two models which estimate the RM evolution over time 
within an SNR. 
First is a model from \citet{2018ApJ...861..150P} which estimates
both the RM and DM evolution for three different scenarios: 
a supernova expanding into a constant density ISM,
a progenitor wind affecting the circumstellar medium,
solely contributing to the RM,
and a supernova expanding into wind affected ISM.
The second model is a one-zone magnetar nebula expanding spherically
at a constant radial velocity \citep{2018ApJ...868L...4M}.
Each model is described in more detail in Appendix \ref{appx:modeldesc}.

Additionally, we consider an environment 
near a massive black hole by comparing to 
the GC
magnetar, \magn{}. 
The RM magnitude and trend of \frb{} seems to be analogous to  
\magn{}, 
which has undergone rapid changes in RM in  
recent times \citep{2018ApJ...852L..12D}.

Using Bayesian inference,
we fit the RM 
evolution prediction from the aforementioned SNR models
to the observed RM of \frb{}.
A Markov--chain Monte Carlo (MCMC) method is used
to estimate 
the posterior of the model parameters
and the age of the \frb{} bursting
source, $t_\mathrm{age}$, at the time of its first
RM measurement.
The models considered here predict that DM decreases over time, 
while the observed DM is increasing. We therefore do not perform
a similar analysis on the DM evolution.
The models do not take into account the short-term stochasticity 
in RM that we observe and our RM uncertainties are also relatively 
small.
We therefore introduce an error added in quadrature in our fitting,
$\Sigma$, to allow for a broader exploration of the 
free parameter space in our MCMC and to obtain more conservtive
uncertainties on the model parameters.
Further details on the fitting can be found in Appendix \ref{appx:modelfit}.
%
%

Henceforth, all values mentioned will be in the
reference frame of the source,
unless otherwise stated. 
This requires a conversion of the observed values 
to the source frame. 
The conversions are $\mathrm{DM}_\mathrm{source}=\mathrm{DM}_\mathrm{obs} (1+z)$,
$\mathrm{RM}_\mathrm{source}=\mathrm{RM}_\mathrm{obs} (1+z)^2$,
and $t_\mathrm{source}=t_\mathrm{obs}(1+z)^{-1}$,
where $z\sim0.2$ is the redshift of \frb{}.
This means that the minimum $t_\mathrm{age}$ possible
in the source frame 
at the time of the first RM measurement
is just over 3 years due to the
time elapsed from the first detection of \frb{} \citep{2014ApJ...790..101S}
and its first RM measurement \citep{2018Natur.553..182M}.
In the case of DM, we will only consider the 
contribution local to the source, i. e. local to the
bursting source and the host galaxy.

\subsection{\citet{2018ApJ...861..150P}} \label{sect:PG}
For the constant density ISM model variety the number
density of the ISM, $n$, is a free parameter,
and for the progenitor wind model varieties the 
wind mass loading parameter, $K$, is a free paramter. 
For all model varieties we also set
$t_\mathrm{age}$ and $\Sigma$ as free paramters
for SN ejecta masses of 10 and 2 $\mathrm{M}_\odot$.
In all cases the SN explosion energy is $10^{51}$~erg.
We estimate the posterier of
$n$, $K$, $t_\mathrm{age}$, and $\Sigma$ for each model variety
\citep[][Eqs. 26, 57, and Appendix]{2018ApJ...861..150P}
using the measured RM values of \frb{} (Table \ref{tab:bursts}). 
Our initial guesses are
the median values of $n$ (1~cm$^{-3}$) and 
$K$ ($10^{13}$~g~cm$^{-1}$) from \citet{2018ApJ...861..150P},
$t_\mathrm{age}=5$~years, and $\Sigma=10^3$~rad~m$^{-2}$ (roughly 1\% of the observed
RM magnitude).
Our results are listed in Table~\ref{tab:PG_models}.

\begin{center}
\begin{table*}
    \begin{center}
    \caption{Model parameters of \citet{2018ApJ...861..150P} for each scenario.
            \textit{From left to right}:
            Model scenario, supernova explosion energy ($E$), 
            supernova ejecta mass ($M$), 
            number density of surrounding uniform ISM ($n$),
            wind mass loading parameter ($K$),
            age of bursting source ($t_\mathrm{age}$),
            and the factor that takes into account additional errors and
            astrophysical variance in RM, $\Sigma$.
            The parameters $n$, $K$, $t_\mathrm{age}$, and $\Sigma$
            were obtained in this work (\S\ref{sect:PG}).
            Uncertainties are 1-sigma.}
    \label{tab:PG_models}
    \begin{tabular}{c c c c c c c}
    Model & E~(erg) & $M$~(M$_\odot$) & $n$~(cm$^{-3}$) & $\log_{10}(K)$~(g~cm$^{-2}$) & $t_\mathrm{age}$~(years) & $\log_{10}(\Sigma$)~(rad~m$^{-2}$) \\
    \hline
    \multirow{2}{*}{Const. ISM} & $10^{51}$ & 10 & $1.7^{+0.1}_{-0.1}$ 
                                & - & $1.4^{+0.2}_{-0.2}$ & $3.9^{+0.1}_{-0.1}$ \\
                                & $10^{51}$ & 2 & $2.5^{+0.1}_{-0.1}$
                                & - & $1.4^{+0.2}_{-0.2}$ & $3.8^{+0.1}_{-0.1}$ \\ \hline
    \multirow{2}{*}{Wind}       & $10^{51}$ & 10 & - & $15.3^{+0.1}_{-0.1}$ 
                                & $7.8^{+0.9}_{-1.1}$ & $3.9^{+0.1}_{-0.1}$\\
                                & $10^{51}$ & 2 & - & $15.6^{+0.1}_{-0.1}$ 
                                & $6.4^{+0.6}_{-0.7}$ & $3.9^{+0.1}_{-0.1}$ \\ \hline
    \multirow{2}{*}{Wind + SNR} & $10^{51}$ & 10 & - & $11.7^{+0.1}_{-0.1}$ 
                                & $8.3^{+1.0}_{-1.2}$ & $3.9^{+0.1}_{-0.1}$\\
                                & $10^{51}$ & 2 & - & $11.9^{+0.1}_{-0.1}$ 
                                & $8.3^{+1.0}_{-1.1}$ & $3.9^{+0.1}_{-0.1}$
    \end{tabular}
    \end{center}
\end{table*}
\end{center}

\begin{figure*}
    \centering
    \includegraphics[width=\textwidth]{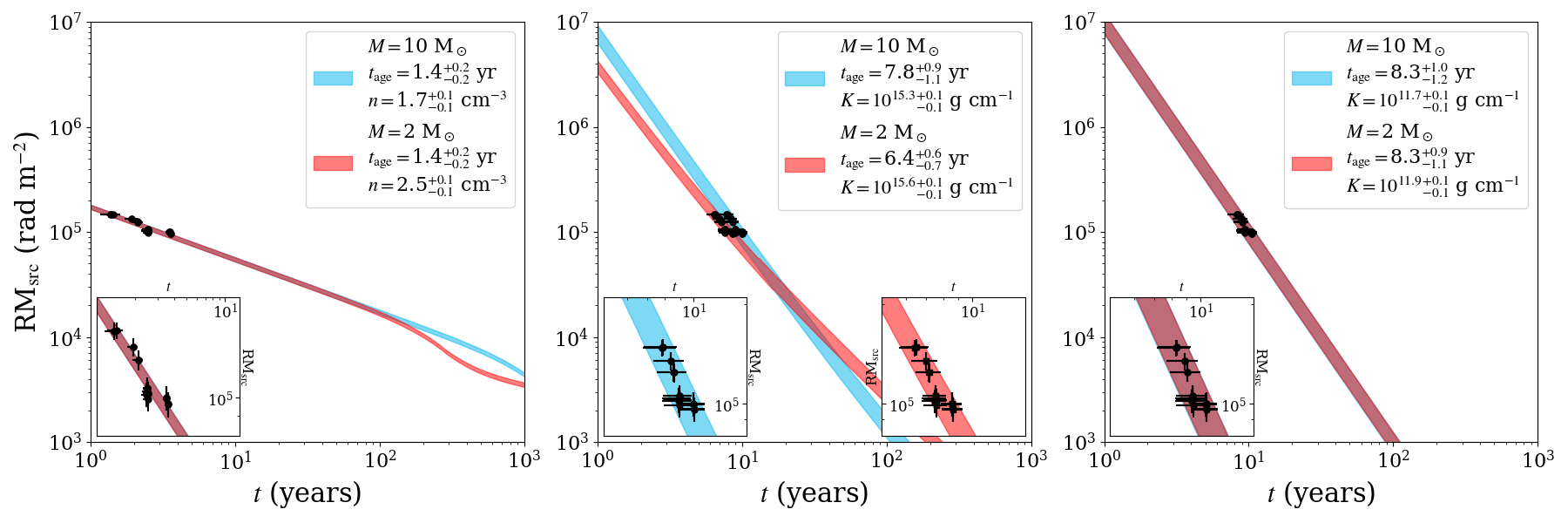}
    \caption{
            Source frame RM as a function of time for each model
            and ejecta mass in \citet{2018ApJ...861..150P}.
            The ranges show the possible RMs from the parameters obtained
            in this work with 1-sigma uncertainties 
            (Table~\ref{tab:PG_models}).
            The black dots are the source frame RMs of \frb{},
            starting at the obtained $t_\mathrm{age}$ for each 
            model variation.
            The RM uncertainties are calculated from Eq.~\ref{eq:sigma}.
            The insets are zoomed in to the RM-time space around
            each $t_\mathrm{age}$.
            \textit{Left}: Uniform ISM model.
            \textit{Center}: Progenitor wind model.
            \textit{Right}: Progenitor wind and evolving supernova 
            remnant model.}
    \label{fig:RM_shade_PG}
\end{figure*}

For the constant ISM model we obtain a 
$t_\mathrm{age}$ of $1.4$~years
at the time of the first RM detection.
For the wind and wind plus SNR evolution models we obtain 
$t_\mathrm{age}$ between $\sim$~6--8~years.
The range of RM from our results (1-sigma error) for each model
and mass is
plotted as a function of time in Fig.~\ref{fig:RM_shade_PG}, and overplotted
with the observed RM values of \frb{}.

\begin{figure}
    \centering
    \includegraphics[width=\columnwidth]{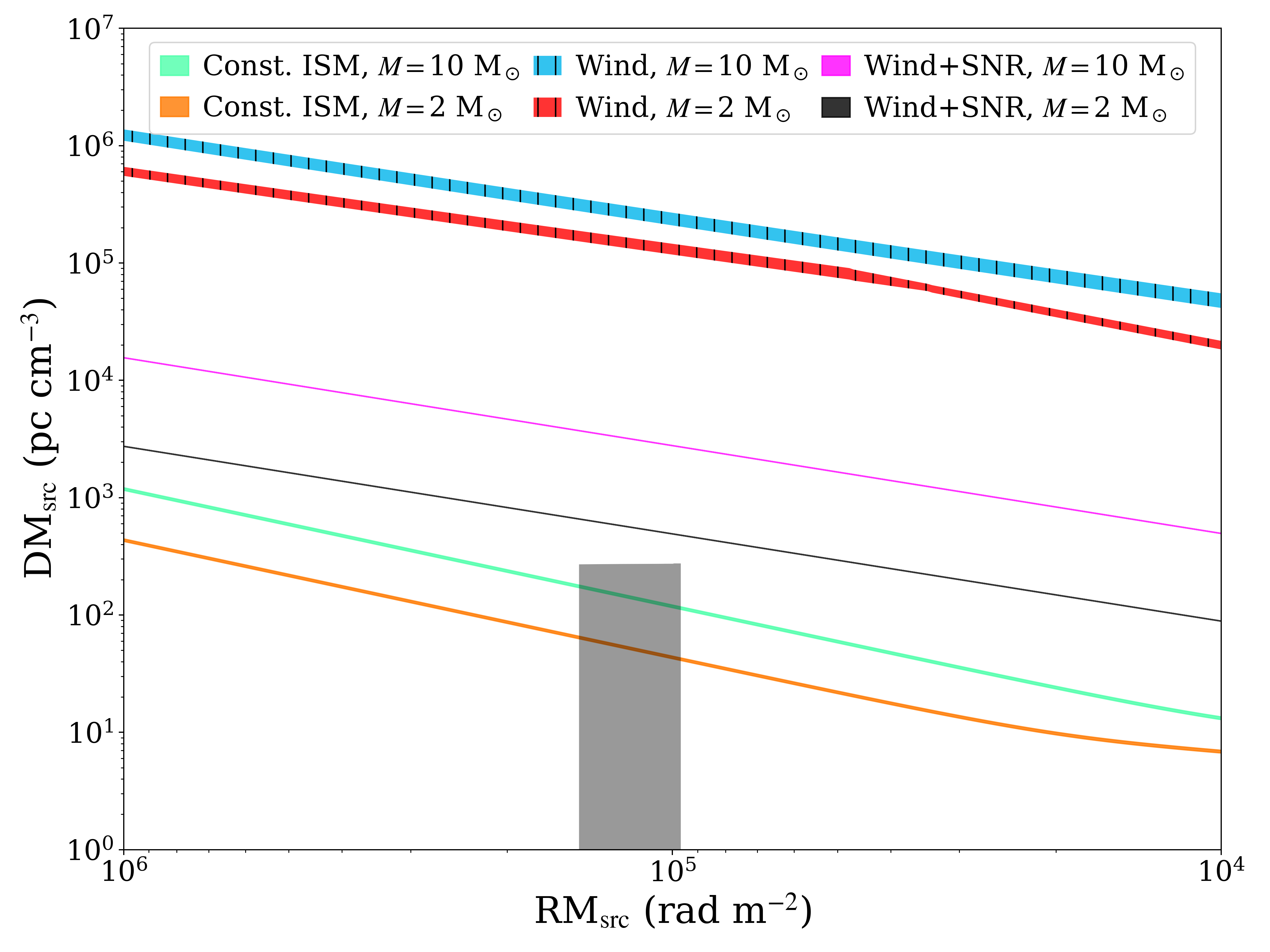}
    \caption{
            Source frame DM versus RM
            of each model scenario presented 
            in \citet{2018ApJ...861..150P}. 
            The width of each line represents the
            range of the predicted
            DMs and RMs of each scenario
            using the parameters obtained in this work
            with 1-sigma uncertainties
            (Table~\ref{tab:PG_models}).
            The grey shaded area shows the RM
            and estimated local source DM contribution of \frb{}
            in the reference frame of the bursting source.
            }
    \label{fig:DMRM_evo}
\end{figure}

We also plot the local DM versus RM for the models in 
\citet{2018ApJ...861..150P} in Fig. \ref{fig:DMRM_evo},
showing how the DM changes as RM decreases over time.
The estimated source frame local DM 
\citep[up to 270~pc~cm$^{-3}$,][]{2017ApJ...834L...7T} and the
source frame RM values of \frb{} are overplotted on
the figure.

\subsection{\citet{2018ApJ...868L...4M}} \label{sect:MM}
In this model we set 
$t_\mathrm{age}$, $\Sigma$, and 
the power-law parameter, $\alpha$, as free parameters.
The free magnetic energy of the magnetar, $E_{B_*}$,
the time since the onset of the magnetar's activity, $t_0$,
and the expansion velocity of the nebula, $v_n$,
differ between model varieties `A, B, and C'.
We estimate the posterior
of $\alpha$, $t_\mathrm{age}$, and $\Sigma$.
The initial guesses are the parameters of
models A, B, and C and $t_\mathrm{age}$ in  
\citet{2018ApJ...868L...4M}, and like before $\Sigma=10^3$~rad~m$^{-2}$.
Our results are listed in Table~\ref{tab:MM_models}.

\begin{center}
\begin{table*}
    \begin{center}
    \caption{
            Model parameters of \citet{2018ApJ...868L...4M}.
            \textit{From left to right:}
            Model, free magnetic energy of the magnetar ($E_{B_*}$), 
            onset of magnetar's active period ($t_0$), 
            radial velocity of expanding nebula ($v_n$), 
            power-law parameter ($\alpha$) and
            age of bursting source ($t_\mathrm{age}$) used in \citet{2018ApJ...868L...4M},
            and $\alpha$, $t_\mathrm{age}$, 
            and the factor that takes into account additional errors and
            astrophysical variance in RM, $\Sigma$,
            obtained in this work (\S\ref{sect:MM}).
            Uncertainties are 1-sigma. 
            }
    \label{tab:MM_models}
    \begin{tabular}{c c c c c c c c c}
    Model & $E_{B_*}$~(erg) & $t_0$~(years) & $v_n$~(cm s$^{-1}$) 
        & $\alpha^\text{a}$ & $t_\mathrm{age}$~(years)$^\text{a}$ 
        & $\alpha^\text{b}$ 
        & $t_\mathrm{age}$~(years)$^\text{b}$ 
        & $\log_{10}(\Sigma)$~(rad~m$^{-2}$)\\
    \hline
    A & $5\times10^{50}$ & 0.2 & $3\times10^8$ & 1.3 & 12.4 
        & $1.6^{+0.3}_{-0.4}$ & $16.8^{+2.0}_{-0.6}$ & $3.9^{+0.1}_{-0.1}$ \\
    B & $5\times10^{50}$ & 0.6 & $10^8$        & 1.3 & 37.8 
        & $1.1^{+0.1}_{-0.1}$ & $16.2^{+2.0}_{-2.6}$ & $3.9^{+0.1}_{-0.1}$ \\
    C & $4.9\times10^{51}$ & 0.2 & $9\times10^8$ & 1.83 & 13.1 
        & $1.6^{+0.3}_{-0.3}$ & $15.3^{+0.8}_{-0.3}$ & $3.9^{+0.1}_{-0.1}$ \\
    \multicolumn{9}{l}{$^\text{a}$\footnotesize{In \citet{2018ApJ...868L...4M}}}\\
    \multicolumn{9}{l}{$^\text{b}$\footnotesize{This work}}
    \end{tabular}
    \end{center}
\end{table*}
\end{center}

\begin{figure*}
    \centering
    \includegraphics[width=\textwidth]{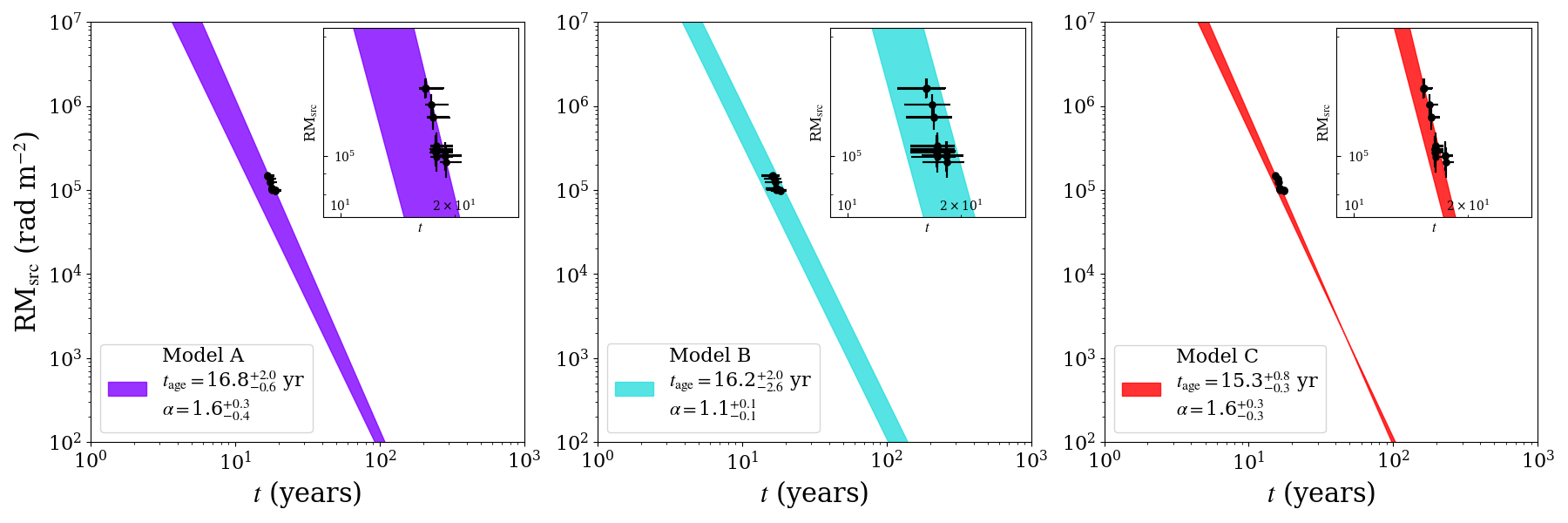}
    \caption{
            Source frame RM as a function of time for each model
            in \citet{2018ApJ...868L...4M}.
            The ranges show the possible RMs from the parameters obtained
            in this work with 1-sigma uncertainties 
            (Table~\ref{tab:MM_models}).
            The black dots are the source frame RMs of \frb{},
            starting at the obtained $t_\mathrm{age}$ for each model.
            The RM uncertainties calculated from Eq.~\ref{eq:sigma}.
            The insets are zoomed in to the RM-time space around each
            $t_\mathrm{age}$.
            \textit{Left}: Model A.
            \textit{Center}: Model B.
            \textit{Right}: Model C.}
    \label{fig:RM_shade_MM}
\end{figure*}

A similar $t_\mathrm{age}$ of $\sim$~15--17~years was obtained
for all the models. 
Our obtained $\alpha$ values lie in the range of 1.1--1.6
and are consistent with the values in \citet{2018ApJ...868L...4M}.
The resulting RM range (1-sigma error),
overplotted with observed \frb{} RM values is
plotted in Fig.~\ref{fig:RM_shade_MM}.

\subsection{Galactic Center Magnetar \magn{}}
The GC magnetar \magn{} has exhibited similar behaviour 
as \frb{} regarding changes in RM.
Since its first RM measurements of $-67000$~rad~m$^{-2}$ 
\citep{2013Natur.501..391E}, it showed some
variations in RM of a few hundred rad~m$^{-2}$ per year 
for a few years until its RM suddenly exhibited a steep
drop in absolute magnitude \citep{2018ApJ...852L..12D}.
This drop in RM is similar to \frb{}, albeit not as intense,
as \magn{} had a drop of 5\% in RM over the course of a year
while the RM of \frb{} has dropped by an average of $15\%$~yr$^{-1}$
over roughly two years.
Both \magn{} and \frb{} exhibit short-term variations
in their observed RMs. Although somewhat similar, the
magnitude of the \frb{} variations is greater.
\citet{2018ApJ...852L..12D} also report a constant DM and
attribute the RM evolution to the changing line of sight towards the
moving magnetar where either the projected magnetic field or
the GC free electron content varies.

\citet{2018ApJ...852L..12D}
use the measured proper motion of \magn{} to estimate 
the characteristic size of magneto-ionic fluctuations to be 
$\sim2$ astronomical units (AU). Assuming the bursts from 
\frb{} originate from the magnetosphere of a neutron star 
with a  
speed of $\sim100$~km~s$^{-1}$, the source moves 
a distance of 20 AU per year.  
The observations of \magn{} show that spatial variations 
on the scale of a few to 10s of AUs are possible in the vicinity 
of a massive black hole. If the host of \frb{} also harbors 
a massive black hole, the variations seen in the RM of \frb{} 
could be caused by the changing medium in its accretion disk. 
The velocity of the medium could be much higher than
in the Galactic center, contributing to the observed fluctuation.

\section{Discussion}
\label{sect:disc}

We compared our measured RM sample to the theoretical RM predictions
of \citet{2018ApJ...861..150P} and \citet{2018ApJ...868L...4M}
by obtaining MCMC posteriors of the model parameters and
the age of the \frb{} bursting source at the time of its
first RM measurement, $t_\mathrm{age}$.

For the model variations in \citet{2018ApJ...861..150P}, we obtain
a $t_\mathrm{age}\sim1.5$~years for the uniform ISM scenario,
and $6-9$~years for the progenitor wind and progenitor wind
plus SNR evolution scenarios.
Based on observations, the minimum possible $t_\mathrm{age}$
is $\gtrsim3$~years, so we exclude the uniform ISM scenario.
A drawback for the wind-only scenario is that it requires a high wind
mass loading parameter ($K>10^{15}$~g~cm$^{-1}$) to be consistent with the data.

We also compare our sample to the predicted DM vs RM evolution in
\citet{2018ApJ...861..150P}. 
The excluded uniform ISM scenario predicts DM
values consistent with \frb{}, 
but both wind scenarios predict much higher DMs than is observed.
However, all the model variations predict a decrease rather than 
increase in DM at the observed source frame RMs of \frb{}. 

Our results here show that origin scenarios with
standard supernovae have difficulties explaining
both the RM and DM of \frb{}.
A caveat is that the models assume uniform media,
while the ISM, SNR, and wind environments most likely have
spatial structures such as filaments.

For the models in \citet{2018ApJ...868L...4M} we obtain a 
$t_\mathrm{age}$ of $\sim$15--17~years
and $\alpha$ of 1.1--1.6.
Our results show that the observed RM evolution
of \frb{} is consistent with these models.
The estimated DM contribution from the nebula 
in \citet{2018ApJ...868L...4M} is $\sim2-20$ pc~cm$^{-3}$
for models A, B, and C (Eq.~\ref{eq:MM_DM}).
The measured increase in DM of $\sim$4 pc~cm$^{-3}$ is difficult
to reconcile with the RM decrease if it originates from the same electrons.

An increase in DM means an increase in the LoS electron density.
There are many contributing factors to the DM along the LoS,
so a smaller fractional change in DM is not surprising.
The Faraday rotating medium contributes only 
a fraction of the total DM and its amount is unknown.
A decrease in RM implies a decrease in the magnetic field
strength or the electron density along the LoS,
or a change in the magnetic field direction.
The opposing RM and DM evolution thus has two possible scenarios:
the changes in RM and DM arise from different media;
or the changes arise from the same medium,
implying that the LoS magnetic field is decreasing in strength
or changing direction.

The DM and RM might not necessarily be coupled.
\citet{2019MNRAS.485.4091M} estimate that photoionization just
outside the propagating outward shock could contribute on the order
of 10 pc~cm$^{-3}$ with an increase of  
a few pc~cm$^{-3}$ possible over several years.
Therefore, the RM decrease and DM increase are likely
occurring in different regions.

The SNR is initially optically thick at radio frequencies due
to free-free absorption.
According to \citet{2016ApJ...824L..32P}, the SNR becomes optically thin at
radio frequencies on a timescale of centuries if the SNR
is solely ionised by the reverse shock.
However, if the SNR is also photoionised from within by the
magnetar wind nebula the SNR becomes optically thin at our observed
frequencies on a timescale of $\lesssim10$~years \citep{2017ApJ...841...14M}.


A by-product of our MCMC calculations is the error
added in quadrature, $\Sigma$. We find $\Sigma$ to be
consistent with $\sim10^{3.9}$~rad~m$^{-2}$
for all models and their variations, or roughly
10\% of the observed RMs. 
The large value of $\Sigma$ compared to the measured
RM uncertainties could be due to deviations of the
observed RMs from the RM evolution models considered
in this work, which are inherently power-laws. 
These deviations could be due to LoS variations
across observing epochs as is seen for \magn{} \citep{2018ApJ...852L..12D},
or due to circumsource turbulence that could account for the
observed stochastic and non-monotonic RM variations. 

An example of RM variations from a source within an SNR is
the Vela pulsar.
Its RM has increased from $34$~rad~m$^{-2}$
up to $46$~rad~m$^{-2}$ ($35\%$ increase) and back down 
to $31.4$~rad~m$^{-2}$ ($32\%$ decrease) between
1970 and 2004 \citep{1985MNRAS.214P...5H,2005MNRAS.364.1397J}. 
Our estimated age of \frb{} is roughly three orders of magnitude
less than Vela, resulting in higher turbulence and denser physical
structures that could cause our observed RM variations.

\magn{} has exhibited similarly drastic changes as \frb{} in
RM over time \citep{2018ApJ...852L..12D}. This change is attributed to 
variations in the projected magnetic field or
the GC free electron content due to line of sight
changes of the moving magnetar.
\frb{} is located outside of its host dwarf galaxy center 
\citep{2017ApJ...834L...7T},
but we cannot exclude a similar scenario due to
the fact that AGNs can be found offset from the 
optical center of dwarf galaxies \citep{2020ApJ...888...36R}.

A comparison can be made between \frb{} and another localised,
repeating FRB, FRB 180916.J1058+65,
which has no discernable 
associated persistent radio source, and its RM is three orders of magnitude
less than the RM of \frb{}. 
However, it can still fit within the SNR framework
where the persistent radio source has faded and the
RM dropped to its observed levels due to the source
being a few hundred years old \citep{2020Natur.577..190M}.

The observed RMs of \frb{}
show large-scale variations of 
$\sim10^4$~rad~m$^{-2}$ over year-timescales
and small-scale variations of $\sim10^3$~rad~m$^{-2}$ 
over week-timescales. 
For a neutron star-black hole system, the 
``cosmic comb'' model \citep{2018ApJ...854L..21Z} predicts a periodic
RM variation correlated with the orbital period of
the neutron star. Such periodicities are not readily
apparent in our data.
There is also no obvious periodicity in the observed RM
variations at the proposed \frb{} periodicity
of 161 days \citep{2020arXiv200803461C}.

Future polarisation measurements will show whether the RM of \frb{} has ``leveled-off'' at its current magnitude
or will continue to vary.
If the RM continues to decrease, the parameters of the SNR models
considered in this work can be constrained further.
On the other hand, if the RM will stay the same,
the models can be rejected or will require adjustments. If the RM increases significantly, it would strongly challenge the SNR models.

Investigating the RM and DM evolution of repeating FRBs
is certainly helpful in constraining source models.
If FRBs, especially repeating ones, continue exhibiting
vast differences from \frb{}, such as host galaxy type, RM magnitude,
and DM evolution, one must consider the possibility that
\frb{} is a unique FRB source, likely residing locally to
an AGN.

\section{Conclusions}
\label{sect:conc}

We present sixteen new RMs from bursts of \frb{}
using observations taken with Arecibo, Effelsberg, and VLA. 

Our Effelsberg survey consists of over 100 observing hours
spanning over two years at $4-8$~GHz (Table~\ref{tab:surveys}). 
An \frb{} survey of this magnitude in this frequency range
is unprecedented, and thus enables us to present a robust,
long-term average burst rate of $0.21^{+0.49}_{-0.18}$~bursts/day
above a fluence of $0.04 \; (w/\mathrm{ms})^{1/2}$~Jy~ms.

Along with previously reported RM values of
\frb{} \citep{2018Natur.553..182M,2018ApJ...863....2G}, 
we have an RM sample spanning roughly 2.5 years.
During that time, the source frame RM has decreased
significantly.
From the first RM measurement at MJD 57747 
to MJD 58215, the RM declined rapidly 
from $1.4\times10^5$~rad~m$^{-2}$ to $1.0\times10^5$~rad~m$^{-2}$.
From that point onward, the RM has stayed relatively
constant, with only a slight decrease down to  
$9.7\times10^4$~rad~m$^{-2}$.
However, short-term RM
variations of $\sim1000$~rad~m~$^{-2}$ per week have
been observed during that period.

We fit the observed RM of \frb{} to theoretical models
of RM evolution from within SNRs
from \citet{2018ApJ...861..150P} and \citet{2018ApJ...868L...4M}.
The results yield a source age estimate of 6--17~years
for \frb{} at the time of its first RM measurement in late 2016. 
Conventional SNRs do not agree with our data, 
but the inclusion of a pulsar wind nebula is
compatible with our data.

\acknowledgments

GHH thanks Dr. Ben Margalit for useful discussions and Dr.~Nataliya~K.~Porayko for help with the MCMC implementation.
The Arecibo Observatory is a facility of the National Science Foundation operated under cooperative agreement (\#AST-1744119) by the University of Central Florida in alliance with Universidad Ana G. Méndez (UAGM) and Yang Enterprises (YEI), Inc.
Based on observations with the 100-m telescope of the MPIfR (Max-Planck-Institut f\"{u}r Radioastronomie) at Effelsberg.
The National Radio Astronomy Observatory is a facility of the National Science Foundation operated under cooperative agreement by Associated Universities, Inc.
DM is a Banting Fellow. LGS is a Lise Meitner Independent Research Group leader and acknowledges support from the Max Planck Society. 
SH acknowledges financial support from the 
Deutsche Forschungsgemeinschaft (DFG) under grant BR2026/25.
RSW acknowledges financial support by the European Research Council 
(ERC) for the ERC Synergy Grant BlackHoleCam under contract no. 610058. 
R.M. recognizes support from the Queen Elizabeth II Graduate Scholarship and the Lachlan Gilchrist Fellowship. 
J.W.T.H. acknowledges funding from an NWO Vici grant (``AstroFlash'').
This publication has received funding from the European Union's Horizon 2020
research and innovation programme under grant agreement No 730562 [RadioNet].
We thank the anonymous referee for their insightful comments.

%

\vspace{5mm}
\facilities{Arecibo, Effelsberg, VLA}


\software{Presto \citep{2011ascl.soft07017R},
            psrfits\_utils,
            PSRchive \citep{2004PASA...21..302H},
            RMsyn.py,
            Single-pulse searcher \citep{2018MNRAS.480.3457M},
            DSPSR \citep{2011PASA...28....1V},
            RM-tools,
            DM\_phase \citep{2019ascl.soft10004S},
            emcee \citep{2013PASP..125..306F},
            scipy \citep{2020SciPy-NMeth}
          }

\pagebreak
\appendix

\section{Model Descriptions}\label{appx:modeldesc}

\subsection{\citet{2018ApJ...861..150P}}
\citet{2018ApJ...861..150P} model the temporal evolution of both RM and DM
of an expanding SNR. 
They consider three cases of evolutionary environments,
which we expand upon below. 

The first evolutionary case is an SNR that expands into 
an ISM of constant density.
The shocked, ionized regions of the SN ejecta and ISM,
as well as ionized material from the pulsar wind nebula
close to the SNR center,
provide sufficient free electrons to disperse an FRB.
The Faraday rotation arises from the magnetic fields 
generated by the forward and reverse shocks during the 
SNR expansion.
The SNR dominates both the DM and RM contributions at
early times until the ISM takes over on a timescale of
$\sim 10^2-10^3$~years.
The free parameters in this model are the number density
of the uniform ISM, $n$, and the SN ejecta mass, $M$.
The energy of the explosion is kept constant as
$E=10^{51}$~erg for all cases.

The second case is where the stellar wind of the massive progenitor affects
the circumstellar environment.
The magnetized wind provides another source of magnetic
field as well as altering the DM evolution.
The DM is much higher initially compared to the previous
scenario due to high density for the wind adjacent to the SN,
but the DM decreases more rapidly because of the wind's
decreasing density.
The wind environment can produce an ordered magnetic field,
which is swept up by the SNR. This is the focal point of
RM generation in this scenario as opposed to the shock
generation of magnetic fields in the previous scenario.
The RM also drops rapidly due to the steep decline with time
of the wind's density and magnetic field.
Here the free parameters are the ejecta mass, $M$,
and the wind mass loading parameter, $K$,
which is a function of the mass loss rate, $\dot{M}$,
and wind velocity, $v_w$, and is given in units of g~cm$^{-1}$.

The third is a mixture of the first two scenarios;
an SNR expands into an ISM affected by a constant velocity wind,
with $M$ and $K$ as free parameters.
For all three cases they assume
supernova ejecta masses of
10~M$_\odot$ (red supergiant progenitor)
and 2~M$_\odot$ (stripped-envelope SN).

\subsection{\citet{2018ApJ...868L...4M}}

\citet{2018ApJ...868L...4M} consider a magnetar
surrounded by a magnetar nebula.
Flares and winds from the magnetar inject particles
and magnetic energy into the nebula that is in turn
responsible for the large observed RM.
Their model is a one-zone magnetar nebula model,
where they assume a spherical, freely expanding nebula with a
constant radial velocity, $v_n$.
The free magnetic energy of the magnetar, $E_{B_*}$ is released
into the nebula at a rate following a power-law in time, $\dot{E}\propto t^{-\alpha}$
\citep[][Eq. 4]{2018ApJ...868L...4M}, where $\alpha\gtrsim1$.
The Faraday rotation occurs in  non-relativistic electrons ejected earlier in the nebula's history and cooled from radiation and adiabatic expansion.

In this model, the RM can be approximated as
\citep[][Eq. 19, values normalised to 1 are omitted for clarity]{2018ApJ...868L...4M}
\begin{equation}
\label{eq:MM_RM}
\begin{split}
    \mathrm{RM}_5 &\approx 6 \; \left(\frac{E_{B_*}}{10^{50}\mathrm{~erg}}\right)^{3/2}
        \left( \frac{v_n}{10^{17}\mathrm{~cm/s}} \right)^{-7/2} \\
        &\times \left( \alpha -1\right)^{3/2} t_0^{(\alpha - 1)/2}
        t^{-(6+\alpha)/2} \mathrm{~rad~m}^{-2},
\end{split}
\end{equation}
where $\mathrm{RM}_5 \equiv \mathrm{RM}/10^5$~rad~m$^{-2}$,
$E_{B_*}$ is in erg, $v_n$ in cm~s$^{-1}$,
$t$ is seconds since the SN explosion,
and $t_0$ is the time in seconds since the onset of the active
period of the magnetar's energy release into the nebula.
We obtain $t_\mathrm{age}$ from Eq.~\ref{eq:MM_RM} by replacing
$t$ with $t_\mathrm{age}+t'$, where $t'$ is the time elapsed in seconds of
each RM measurement since the first one.

For completeness, the estimated DM contribution from the Faraday-rotating medium is given by
\begin{equation}
\begin{split}
    \mathrm{DM}&\sim3\times10^{18} \; \left(\frac{E_{B_*}}{10^{50}}\right) 
        \left(\frac{v_n}{10^8}\right)^{-2} \\
        &\times (\alpha-1) t^{-2}
        \mathrm{~pc~cm}^{-3}
\label{eq:MM_DM}
\end{split}
\end{equation}

In their analysis, \citet{2018ApJ...868L...4M} consider three
variations of their model with each having its own set of
values for $E_{B_*}$, $t_0$, $v_n$, and $\alpha$.
They call these variations `model A, B, and C', and
we keep the same notation to avoid confusion.
\citet{2018ApJ...868L...4M} use models A, B, and C to estimate
$t_\mathrm{age}$ of \frb{} from Eq.~\ref{eq:MM_RM}
using the RM measurements from \citet{2018Natur.553..182M}
and \citet{2018ApJ...863....2G}.
Their choice of parameters and their results
are shown in Table \ref{tab:MM_models}.

\section{Model Fitting}\label{appx:modelfit}

To perform an MCMC we used the 
\texttt{emcee}\footnote{\url{emcee.readthedocs.io}} Python package \citep{2013PASP..125..306F}.
MCMC deploys random walkers around the initial estimates
of the parameters,
where the walkers explore the parameter space
in order to reconstruct the posterior probability of
the parameters.

To obtain an initial estimate for our parameters we used the \texttt{scipy}
\citep{2020SciPy-NMeth}
stochastic least squares module \texttt{differential\_evolution}.
An initial guess is also required for \texttt{differential\_evolution},
where we used the parameters of each model variety in
\citet{2018ApJ...861..150P} and \citet{2018ApJ...868L...4M}.
For our MCMC we randomly scattered 10 walkers around each parameter
(up to 10\% away), where each walker was made to walk $1.5\times10^3$ steps.
We used uninformative uniform priors for all our model parameters.

We introduced an error added in quadrature, $\Sigma$, 
to our observed RM uncertainties
to be able to properly explore the free parameter space
and to obtain more conservative uncertainties on our model parameters.
$\Sigma$ enters our Gaussian likelihood function as an
underestimation factor of the variance $\sigma$ (observed RM uncertainties
in this case) as
\begin{equation}
\label{eq:sigma}
    s^2 = \sigma^2 + \Sigma^2.
\end{equation}
The measured RM uncertainties, $\sigma$, are on the order of $\lesssim$~$10^2$~rad~m$^{-2}$.

\begin{figure*}
    \centering
    \includegraphics[width=.4\textwidth]{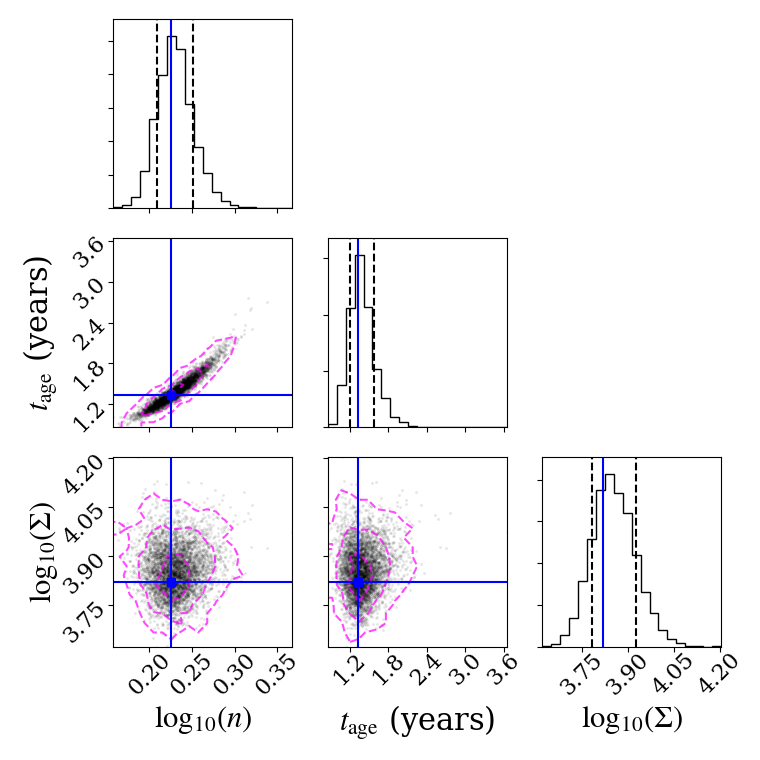}
    \includegraphics[width=.4\textwidth]{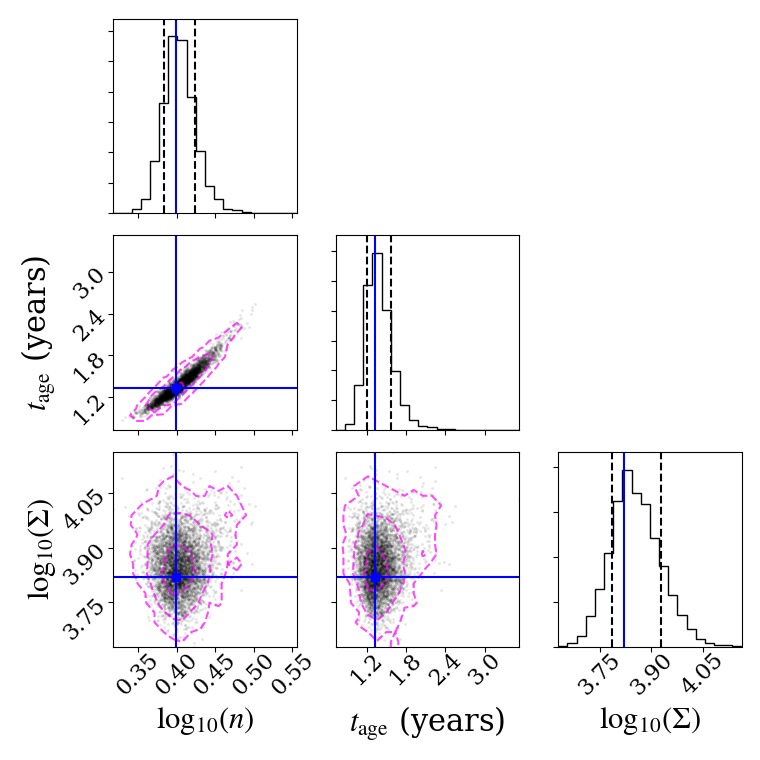}
    \includegraphics[width=.4\textwidth]{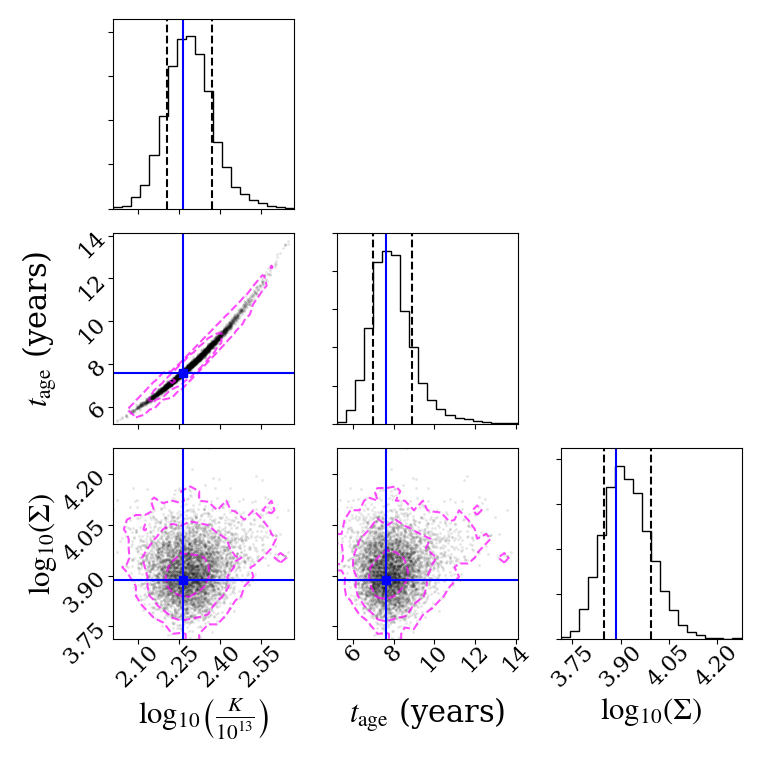}
    \includegraphics[width=.4\textwidth]{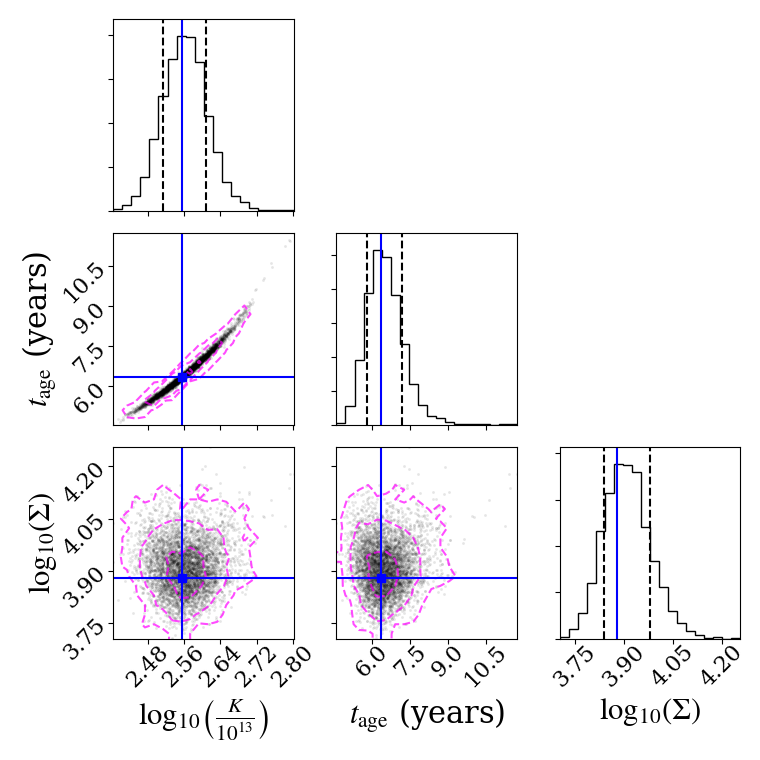}
    \includegraphics[width=.4\textwidth]{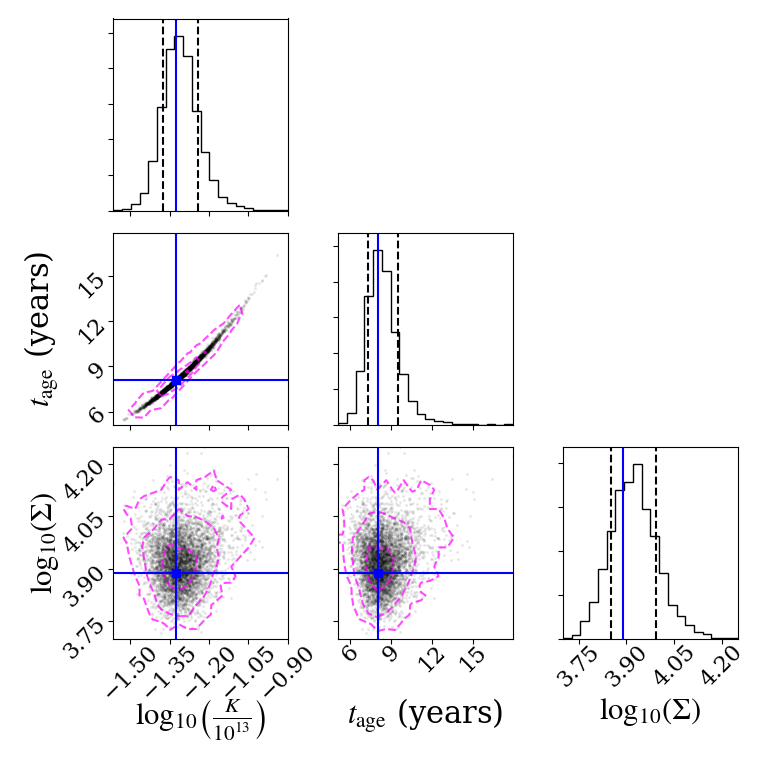}
    \includegraphics[width=.4\textwidth]{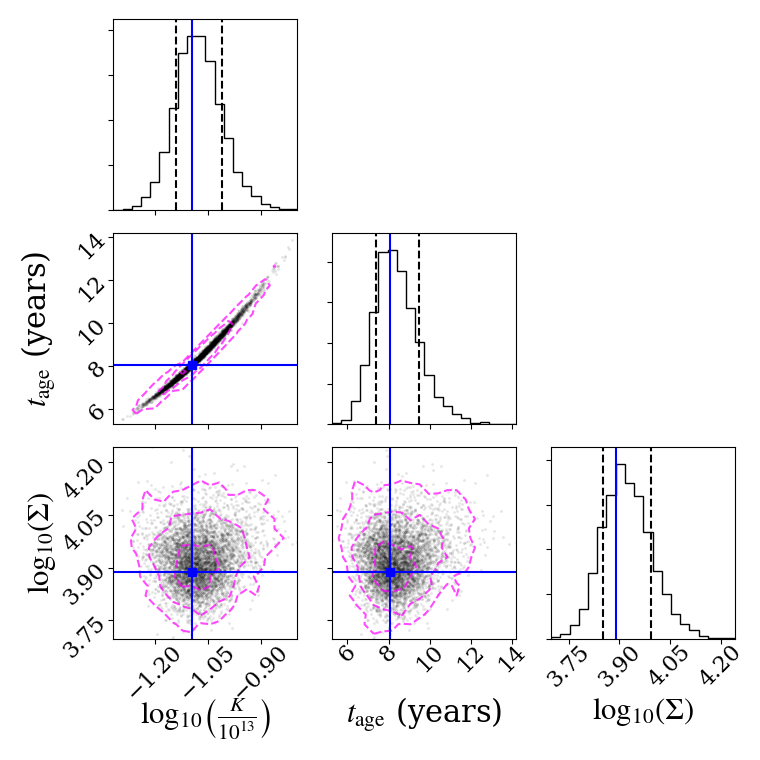}
    \caption{2D posterior corner plot for the parameters 
            ($n$, $K$, $t_\mathrm{age}$, and $\Sigma$) 
            of the models in \citet{2018ApJ...861..150P}.
            The histograms indicate the posterior probability
            of each parameter, with the dashed vertical lines denoting
            the 1-sigma range. 
            The plots show the explored parameter space, with
            1, 2, and 3 sigma dashed contours.
            The crosses indicate the prior used for each parameter,
            obtained with a stochastic least squares method.
            The left column shows the results for a 10~M$_\odot$ ejecta,
            and the right column for a 2~M$_\odot$ ejecta.
            \textit{Top row:} Uniform ISM model.
            \textit{Middle row:} Progenitor wind model.
            \textit{Bottom row:} Progenitor wind and evolving 
            supernova remnant model.
             }
    \label{fig:PG_corner}
\end{figure*}

\begin{figure*}
    \centering
    \includegraphics[width=.4\textwidth]{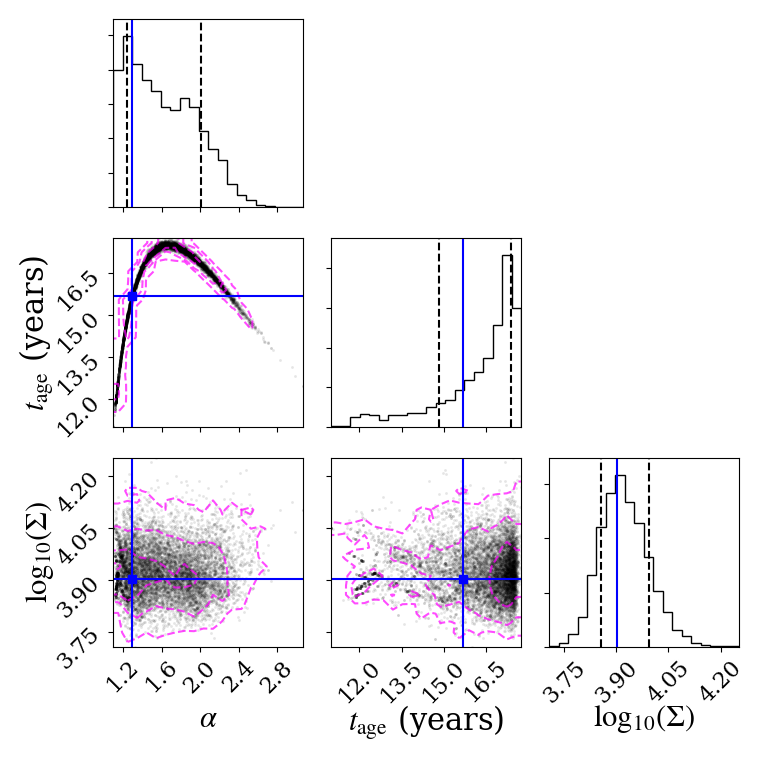}
    \includegraphics[width=.4\textwidth]{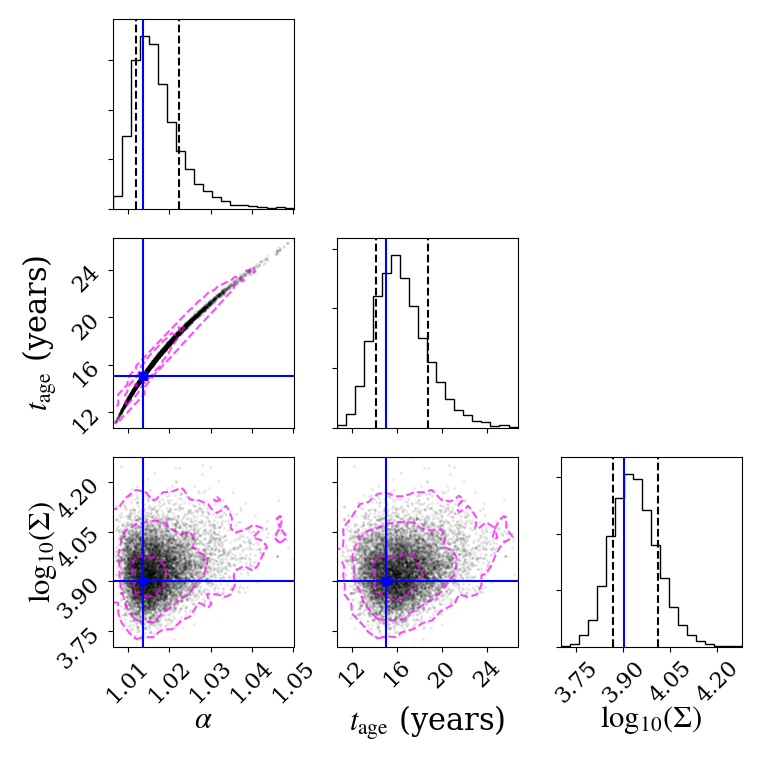}
    \includegraphics[width=.4\textwidth]{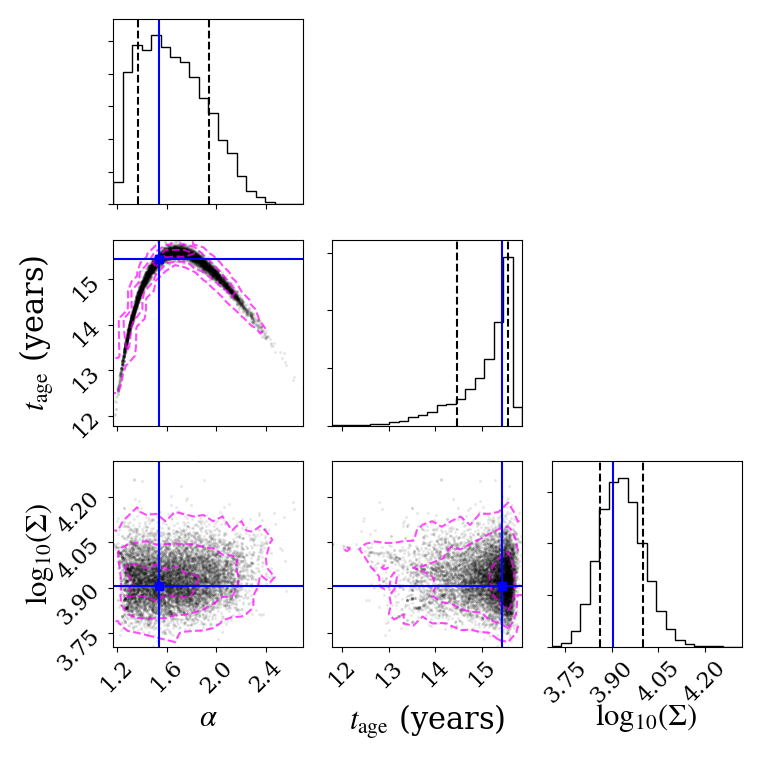}
    \caption{2D posterior corner plot for the parameters 
            ($\alpha$, $t_\mathrm{age}$, and $\Sigma$) 
            of the models in \citet{2018ApJ...868L...4M}.
            The histograms indicate the posterior probability
            of each parameter, with the dashed vertical lines denoting
            the 1-sigma range. 
            The plots show the explored parameter space, with
            1, 2, and 3 sigma dashed contours.
            The crosses indicate the prior used for each parameter,
            obtained with a stochastic least squares method.
            \textit{Top left:} Model A.
            \textit{Top right:} Model B.
            \textit{Bottom:} Model C.
             }
    \label{fig:MM_corner}
\end{figure*}

The 2D posterior corner plots obtained from fitting the models of
\citet{2018ApJ...861..150P} and 
\citet{2018ApJ...868L...4M} to our data are shown in 
Figs.~\ref{fig:PG_corner}--\ref{fig:MM_corner}.

\bibliography{bibby}{}
\bibliographystyle{aasjournal}



\end{document}